\documentclass[journal]{IEEEtran}
\pdfoutput=1
\UseRawInputEncoding
\usepackage{multirow}
\usepackage{diagbox}
\usepackage{amssymb}
\usepackage{amsmath}
\usepackage{graphicx}
\usepackage{cite}
\usepackage{citesort}
\usepackage{subfigure}
\usepackage{graphicx,epstopdf}
\usepackage{epsfig}	
\usepackage{cite,graphicx,amsmath,amssymb}
\usepackage{comment}
\usepackage{algorithm}
\usepackage{algorithmic}
\usepackage{graphicx}

\usepackage{amssymb}
\usepackage{amsmath}
\usepackage{cite}
\usepackage{url}
\usepackage{xcolor}
\usepackage{cite,graphicx,amsmath,amssymb}
\usepackage{subfigure}
\usepackage{citesort}
\usepackage{fancyhdr}
\usepackage{mdwmath}
\usepackage{mdwtab}
\usepackage{caption}
\usepackage{amsthm}
\usepackage{lipsum}

\newtheorem{theorem}{Theorem}

\newtheorem{lemma}{Lemma}

\newtheorem{corollary}{Corollary}

\newtheorem{remark}{Remark}  

\makeatletter
\def\ScaleIfNeeded{%
\ifdim\Gin@nat@width>\linewidth \linewidth \else \Gin@nat@width
\fi } \makeatother

\begin{document}
\title{\Huge{A Unified NOMA Framework in Beam-Hopping Satellite Communication Systems}}

\author{ Xuyang~Zhang,~\IEEEmembership{Xinwei~Yue,~\IEEEmembership{Senior Member~IEEE}, Tian Li,~\IEEEmembership{Member,~IEEE}, Zhihao Han, Yafei Wang, Yong Ding and  Rongke\ Liu,~\IEEEmembership{Senior Member~IEEE} }

\thanks{This work was supported in part by the National Key R\&D Program of China under Grant 2020YFB1807102, in part by National Natural Science Foundation of China under Grant 62071052, Grant 62201533, in part by the Beijing Natural Science Foundation under Grant L202003, and in part by the Civil Aerospace Technology Advance Research Project of the 13th Five-Year Plan under Grant D030301. \emph{(Corresponding author: Xinwei Yue.)}}
\thanks{X. Zhang, X. Yue and Y. Wang are with the Key Laboratory of Information and Communication Systems, Ministry of Information Industry and also with the Key Laboratory of Modern Measurement $\&$ Control Technology, Ministry of Education, Beijing Information Science and Technology University, Beijing 100101, China (email: \{xuyang.zhang, xinwei.yue and wangyafei\}@bistu.edu.cn).}
\thanks{T. Li is with the 54th Research Institute of China Electronics Technology Group Corporation, Shijiazhuang Hebei 050081, China. (email: t.li@ieee.org).}
\thanks{Z. Han and R. Liu are with the School of Electronic and Information Engineering, Beihang University, Beijing 100191, China. R. Liu is also Shenzhen Institution of Beihang University, Shenzhen 518063, China  (email: \{hzh$\_$95, rongke$\_$liu\}@buaa.edu.cn).}
\thanks{Y. Ding is with the Baicells Technologies Co. Ltd, Beijing 100095, China (email: dingyong@baicells.com).}
 }




\maketitle

\begin{abstract}
This paper investigates the application of a unified non-orthogonal multiple access framework in beam hopping (U-NOMA-BH) based satellite communication systems. More specifically, the proposed U-NOMA-BH framework can be applied to code-domain NOMA based BH (CD-NOMA-BH) and power-domain NOMA based BH (PD-NOMA-BH) systems. To satisfy dynamic-uneven traffic demands, we formulate the optimization problem to minimize the square of discrete difference by jointly optimizing power allocation, carrier assignment and beam scheduling. The non-convexity of the objective function and the constraint condition is solved through  Dinkelbach's transform and variable relaxation. As a further development, the closed-from and asymptotic expressions of outage probability are derived for CD/PD-NOMA-BH systems. Based on approximated results, the diversity orders of a pair of users are obtained in detail. In addition, the system throughput of U-NOMA-BH is discussed in delay-limited transmission mode. Numerical results verify that: i) The gap between traffic requests of CD/PD-NOMA-BH systems appears to be more closely compared with orthogonal multiple access based BH (OMA-BH); ii) The CD-NOMA-BH system is capable of providing the enhanced traffic request and capacity provision; and iii) The outage behaviors of CD/PD-NOMA-BH are better than that of OMA-BH.
\end{abstract}
\section{Introduction}
With the seamless coverage of the global information network and the interconnection of all things, the terrestrial wireless network will integrate with satellite communication to form an integrated mobile internet of air-space-ground\cite{Ye2020SAG}. The air-space-ground networks were designed to provide seamless wide-area, high-throughput, and evenly distributed elastic links for the sixth-generation (6G) communication networks\cite{chen2020B5G6G}. Compared with geostationary earth orbit satellites, the primary feature of low earth orbit (LEO) satellites is the lower construction costs, low path loss and latency. With respect to terrestrial communications, the LEO satellite communication systems have ability to meet the requirements of wider coverage and long distance\cite{Gu2022LEOGEO}. To overcome non-uniform traffic demand allocation and maximize the utilization of satellite payload \cite{Abdu2021Res},  the frequency multiplex in accordance with acceptable interference was advanced seriously. In \cite{Fu20216G}, the resource management efficiency of ground-relay-satellite links was enhanced by regulating satellite-relay antennas angle. In addition, the efficient and flexible performance of high-speed satellite-terrestrial links was investigated \cite{Zhang2021edge}, which takes into consideration the deployment of edge computing servers at gateway stations.

Relative to orthogonal multiple access (OMA), non-orthogonal multiple access (NOMA) has more obvious advantages in the light of the fairness, flexibility of scheduling and maximum capacity\cite{Ding2017App,Li2020S}. Concerning avoiding user resource conflict during NOMA transmission, spreading codes were used\cite{C2022OMA}. Driven by this trend, the unified framework of NOMA was proposed, which includes code-domain NOMA (CD-NOMA) and power-domain NOMA (PD-NOMA)\cite{Yue2018Unif}. More particularly, the authors of\cite{Mol2018SCMA} studied the conquering resource waste problem of sparse code multiple access  in heterogeneous cellular networks, where the CD-NOMA based systems achieved a better sum rate at the cost of more complexity. As mentioned in \cite{DE2021OTFS}, a multi-user orthogonal time-frequency-space system with CD-NOMA technology was analyzed to endure the dares of orthogonal frequency division reusing in high Doppler conditions. The authors in \cite{Han2018serial} described a serial code suggestion for system sequence cutting back, which reduces detection complexity compared to OMA and other NOMA schemes. To more raise the system efficiency, the intelligence reflecting surface assisted CD-NOMA was surveyed in \cite{Dai2018PDMA}, where the patterns of multiple users were intelligently designed. Regarding PD-NOMA, the effect of multiple users resource allocation in PD-NOMA was characterized using successive interference cancellation (SIC) in\cite{Li2020}. Moreover, the authors of \cite{Mog2022JOINT} discussed a joint capacity optimization and coordinated storing data program for heterogeneous networks based on PD-NOMA, aiming to minimize the quality of service. From the parse of variable balanced equitable dispatching in downlink PD-NOMA networks \cite{Liu2018fair}, a highly efficient programme was researched by combining power distribution and terminal congregation picking. In addition, the QoS-guarantee node power optimization algorithm in constraints of total satellite power and transmission delay has been investigated \cite{2021Qos}, while the ground industrial internet of things system model was adopted with a NOMA-based Ka-band multibeam satellite. The authors of \cite{2021Ge} proposed NOMA optimization for multilayer satellite networks, where the multi-user pairing problem was transformed into a max-min pairing algorithm to guarantee fairness. A Hungarian joint dichotomous algorithm based on the scheme of paired NOMA users has been discussed in which the satellite calculates the channel condition ratio for resource allocation and users-base station association\cite{2021stsn}.

As one of key technologies for 6G networks, the beam hopping (BH) has ability to drop the transmission delay for multi-beam satellite networks\cite{Zhang2018BH,Shao2021BH}. In actual, the BH was regarded as the allocation of time domain resources on the satellite, where the dwell time of the wave position coverage area was changed according to the traffic needs and satellite load capacity\cite{Cai2022inter}. With the aim of solving satellite packet queuing delay, the authors of \cite{Tang2021BH} employed a novel mode for active beam position regulation of BH satellite communication scheme. In \cite{Lin2022BH}, the authors explored a flexible beam design and bandwidth assignment network of deep reinforcement learning according to three variables of time, space, and frequency. From the perspective of carrier-to-interference ratio, the authors in \cite{Fon2012BH}  evaluated the antenna systems of combining BH and dimension drop of plenty applied speckles. Furthermore, the improved cuckoo algorithm was optimized to service a great deal of high-weight terminals as possible in the BH networks\cite{Zhang2021BH}. In order to reduce operating costs, the authors of \cite{Koosha2019} discussed a blended method for progressive overhead throughput satellite systems utilizing BH. Recently, a innovative joint BH and precoding scheme was surveyed in \cite{Xu2022NOMA}, where the energy consumption optimization and intra-beam interference inhibition were combined. From the perspective of non-uniform traffic allocation \cite{Li2021Benefit}, the authors confirmed the performance advantages of BH compared to the multiple polarization-frequency reuse networks.

For another special case, the coupling of off-the-ground networks and NOMA is capable of achieving regional full coverage and enhance transmission efficiency. The application of NOMA to satellite communications has received a great deal of attention\cite{Yue2020NOMASatellite,Wang2021mis,Gao2022files}. As propounded in \cite{Yue2020NOMASatellite}, the authors investigated the outage probability of NOMA based satellite communications over shadowed-rician fading Channels.
In \cite{Wang2021mis}, the authors employ NOMA to ease intra-beam interference while achieving precoding to decrease inter-beam influence.
A NOMA based alliance conformation gamble medium was proposed for optimizing overall cost\cite{Gao2022files}, when  the satellite transmits data to the principal terminals of each group by the NOMA.
An unmanned aerial vehicle space-ground-air integral relay network was proposed\cite{Wang2022UAV}, where the NOMA-assisted unmanned aerial vehicle was equipped with a phased array antenna to receive satellite signals. According to the advantages of downlink NOMA-BH in LEO satellite systems, the authors of \cite{Gao2020analysis} derived the ergodic capacity, the outage probability and mutual information of two users in a spot beam. As a further achieving non-orthogonality in power and frequency domains of the BH network \cite{Jia2019}, the design of bandwidth compression was embedded in the NOMA scheme. Based on the optimization of BH designed in multiple antenna scenarios\cite{Li2021}, the two-step resource optimization algorithm better raising resource utilization was verified. The authors of\cite{Wang2021joint} investigated minimizing traffic matching errors by jointly optimizing time slots and power allocation. Inspired by this point, the upper-bound approach and lower-bound approach have been  highlighted through an efficient boundary way in\cite{Wang2022Joint}.
\subsection{Motivation and Contributions}
The above research works provide a solid foundation for understanding the performance of U-NOMA-BH systems, the papers for researching the unexcavated advantages by uniting these two progressive technologies are still in the groping stage of the field. As mentioned above, the existing research contributions focus attention NOMA and BH operations on the power domain, which leads to the satellite intra-beam interference and excessive complexity of SIC receivers. However, the CD-NOMA-BH system can accomplish user pairing and isolate intra-beam interference  with limited carrier resources. The PD-NOMA-BH system is designed as a special case continuation of the idea. In addition, it is difficult to obtain perfect channel state information (pCSI) in practical communication scenarios due to the effect of path loss, rain fading and atmospheric ionization. Hence the impact of imperfect channel state information (ipCSI) on the performance of U-NOMA-BH systems should be considered. The authors of \cite{Wang2022Joint} proposed a greedy algorithm to solve the mixed integer non-convex programming (MINCP) for PD-NOMA-based BH systems.  From practical considerations, considering only the PD-NOMA-BH system may lead to the low utilization of satellite carrier resources. To the best of our knowledge, there is no related work to analyze the optimization and performance of unified NOMA based BH (U-NOMA-BH) framework, which prompts us to develop this paper. More specifically, we formulate the optimization of the U-NOMA-BH systems for multiple users satisfaction and propose two corresponding optimization algorithms. Then, we further derive the closed-form expressions of outage probability and system throughput for the CD/PD-NOMA-BH systems with ipCSI/pCSI. The essential contributions of our paper are summarized as follows:
\begin{enumerate}
  \item  We formulate a resource optimization scheme for the satisfaction of users, which minimizes the gap between users achievable capacity and traffic requests, i.e., the paired NOMA users traffic gap composed of users in beam-center users (BCU) and beam-edge users (BEU). On the basis of the proposed problem, the satisfaction of users is constrained by optimizing carrier variables, power allocation factors and beam-level time slots. Furthermore, the optimization problem and constraints are transformed into convex problems for analysis.
  \item We design an efficient joint optimization scheme and develop an iterative optimal algorithm to solve the convex optimization problem. Multi-carrier users are optimized by applying sparse matrix matching. The optimization problem is divided into several subproblems, where the time slot allocation is based on user requirements. The simulation results show that U-NOMA combined with BH has excellent performance.
  \item We derive the closed-form expressions of outage probability for the BCU and BEU by exploiting the U-NOMA-BH optimization algorithm, respectively. For obtaining more insights, we further derive the asymptotic outage probability of two users. Furthermore, we calculate the diversity orders and the system throughput in delay-limited transmission according to the derived outage probability for U-NOMA-BH systems.
  \item We study the outage probability and system throughput of BCU and BEU with ipCSI. The results show that the outage performance of BEU with pCSI is better than OMA. We compare the performance of BCU and BEU under different channel error parameters and overlapping time slots. We observe that U-NOMA-BH systems are capable of providing better fairness in the multi-user state.
\end{enumerate}
 \subsection{Organization and Notations}
 The U-NOMA-BH communication systems are illustrated in Section II. In Section III, we formulate a joint resource optimization problem and elaborate the operational steps for CD-NOMA-BH and PD-NOMA-BH, respectively. In Section IV, we analyze outage probability and system throughput for the closed form solutions of BCU and BEU. Section V provides the numerical results to verify the derived analytical results. In  Section VI concludes the paper.

The main notations of this paper are explained as follows: $f_{X}(\cdot)$ and $F_{X}(\cdot)$ denote the probability density function (PDF) and cumulative distribution function (CDF) of a random variable \emph{X}, respectively; $(\cdot)^T$ represents stand for transpose operations; $\|\cdot\|_{2}^{2}$ denotes Euclidean two norm of a vector; $diag( \cdot )$ represents a diagonal matrix.
\section{System Model}\label{System Model}
We consider a unified NOMA framework to BH communication scenarios, as shown in Fig. 1, where multiple users receive information from an LEO satellite. More precisely, the coverage area of satellite is planned as $B$ beam positions\footnote{It is worth noting that considering time-varying multiple LEOs in the U-NOMA-BH systems can further cover the changing regions, which will be considered in our future work.}, and $\mathcal{B}$ is the set of active beams. Note that the users are randomly assigned to beams according to geographic coordinates. The users are divided into the user \emph{n} of BCU and user \emph{m} of BEU respectively based on channel conditions in each beam, where the user $n$ and user $m$ are combined into paired NOMA users. The gateway collects users traffic demand and CSI from ground users delivering LEO via the feedback link. According to the feedback, the resource manager optimizes the beam illumination pattern based on the resource allocation of algorithms used for U-NOMA-BH systems. It is assumed that each user is equipped with one antenna. The beam of satellite directly covers to $M$ ground users, generating a sparse user-subcarriers matrix  ${{\bf{G}}_{K \times M}}$ in the form of an overload (i.e., the information of user $n$ and user $m$ is mapped to the $K$ sparse subcarriers (SC), which satisfies the relationship of $M > K$)\footnote{As adopted by many multiple access concepts, the CD-NOMA-BH mainly adopts sparse code multiple access (SCMA) or pattern division multiple access (PDMA). Actually, CD-NOMA-BH is considered to be an extension of SCMA.}. One frequency band serves all beams in common, i.e., the mode transmission of single multiplexing\cite{Couble2018Two}.

The application scenarios for the combination of LEO and BH are the areas of non-uniform users' demand and poor ground base stations, such as mountains, disaster relief and non-hot spot area services. Besides, the U-NOMA-BH systems can overcome shadow blocking and accomplish  flexible resource allocation \cite{impact_SCMA}. The BH satellite systems utilize multi-port amplifiers and phased array antenna technology to achieve excellent resource allocation capabilities. The implementation equipments of PD-NOMA-BH system consists of multi-port power amplifier and phased array antenna. For further explanation, the multi-port power amplifier selects the corresponding number of ports and allocates the power of each port by the algorithm used for PD-NOMA-BH system\cite{2007IVEC}. The phased array antenna changes the active beam position of BH by controlling the phase of the radiation unit. Based on these, the codebook is mapped to physical frequency domain resources by the algorithm of the CD-NOMA-BH system. Meanwhile, the codebook adopts high-dimensional quadrature amplitude modulation (QAM) to more easily decode non-orthogonal user signals at the receiver side \cite{2019PAPR_wcsp,2019MPA}. According to ground users feedback, the input signal matrix is regrouped into the output matrix by adjusting the NOMA power allocation of the multi-port amplifier. Based on the on-board payload capability, the BH system adopts the method of time division and space division, and the phased array antenna adjusts the feed phase flexibly to change the beam pattern. Assuming that the satellite is equipped with a digital transparent transponder, the satellite transmits ground data to users as a relay. The ground stations solve queuing problem for beam allocation in the BH systems. The joint allocation problem of time slots, subcarriers and NOMA power factors is optimized through the network resource manager.
\begin{figure}[t!]
    \begin{center}
        \includegraphics[width=3.4in,  height=2.6in]{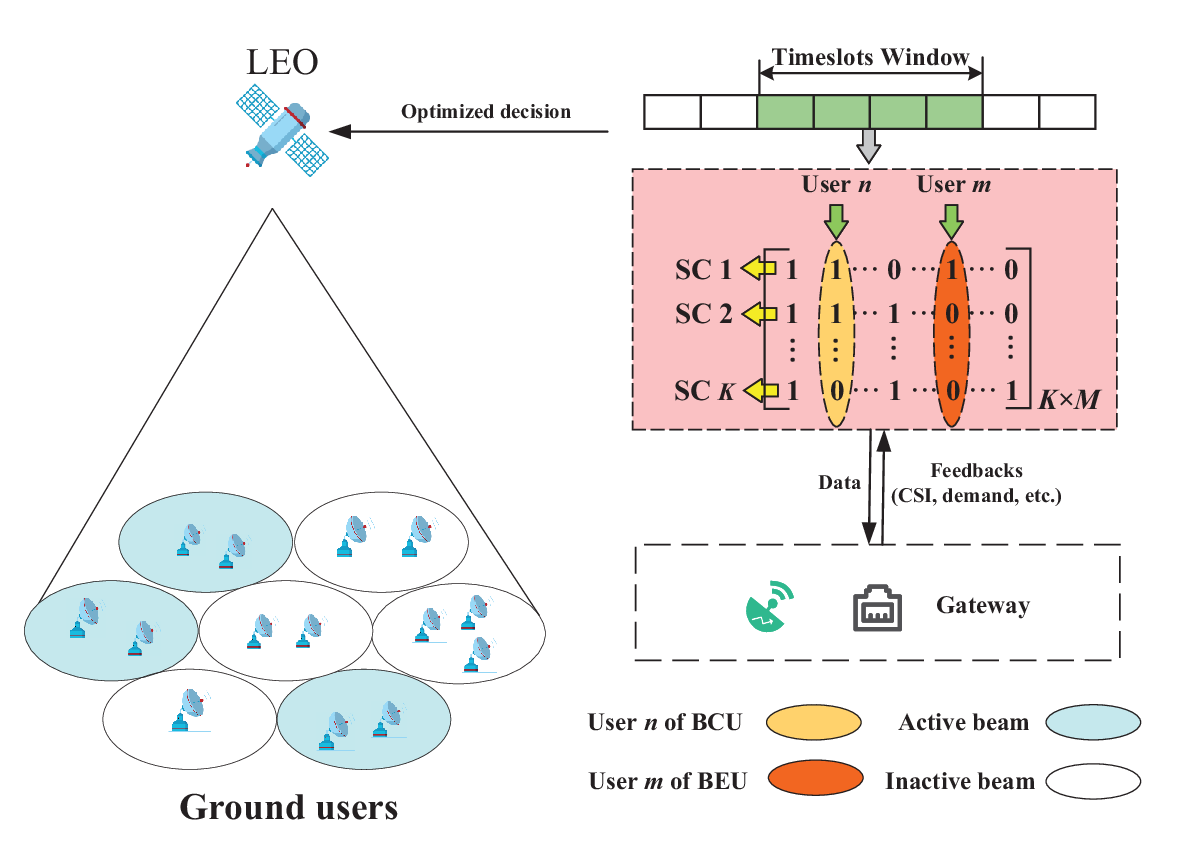}
        \caption{An illustration of U-NOMA-BH satellite communication systems.}
         \label{system model}
    \end{center}
\end{figure}
In CD/PD-NOMA-BH systems, the number of active beams ${B_0}$ in each time slot exceed the total number of beam positions is defined, i.e.,  ${B_0} \le B$. For convenience, the maximum number of active beams is mentioned later with reference to the number of all beam positions  $B$. The scheduling period of the beam defines the BH window, which consists of \emph{T} time slots. ${\cal T}$  represents the set of time slots. Let ${\cal M}$ and ${\cal N}$ represent the set of user \emph{m} and user \emph{n}, respectively. Randomly selecting user $\varphi $, i.e., $\varphi \in(n, m)$ over \emph{t} time slot is given by
 \begin{align}\label{1}
 \mathbf{y}_{\varphi, t}= \nonumber
 &\operatorname{diag}\left(\mathbf{h}_{\varphi, t}\right)\left(\mathbf{g}_{n, t} \sqrt{P_{s} a_{n, t}} x_{n}+\mathbf{g}_{m, t} \sqrt{P_{s} a_{m, t}} x_{m}\right)\\\nonumber
 &+\mathbf{I}_{i n t e r , t}+\mathbf{I}_{i n t r a, t}+\omega^{*} P_{s}+\sigma^{2},\tag{1}
\end{align}
 where ${a_m}$ and ${a_n}$ represent the power allocation factors of user \emph{m} and user \emph{n} in the paired NOMA users, respectively. Considering the fairness of users,  ${a_n}$ should be less than $a_{m}$ and ${{a_m} + {a_n} =1}$. $\sigma^{2}$ represents the Gaussian white noise. Let ${{\bf{h}}_\varphi } = {\left[ {{h_{\varphi 1}}{h_{\varphi 2}} \cdots {h_{\varphi k}}} \right]^T}$  denotes the channel vector of carrier $k$ occupied by user $\varphi$ within a satellite beam. Without loss of generality, the effective channel gains between the satellite and users are ordered as  $\left\| {{{\bf{h}}_{M,t}}} \right\|_2^2 >  \cdots  > \left\| {{{\bf{h}}_{n,t}}} \right\|_2^2 >  \cdots  > \left\| {{{\bf{h}}_{m,t}}} \right\|_2^2 >  \cdots  > \left\| {{{\bf{h}}_{1,t}}} \right\|_2^2$. The users are paired by the carrier distribution requirements, where paired NOMA users are transmitted over the same subcarrier. As a result of the channel estimation error in the actual satellite communication network, the actual channel model can be expressed as ${\bar h_{\varphi k}} = {h_{\varphi k}} + {e_{\varphi k}}$, where $e_{\varphi k} \sim \mathcal{C N}\left(0, w^{*}\right)$. We introduce the widely adopted channel model in satellite systems, which is derived as ${{\ h}_{\varphi k}} = {\eta _{\varphi k}}\sqrt {{G_r}} \sqrt {{{\left( {\frac{\lambda }{{4\pi {d_\varphi }}}} \right)}^2}} \sqrt {{G_t}} $, where $\eta_{\varphi k} \sim \mathcal{C N}(0,1)$ represents the channel coefficient. Additionally, for free space path loss factors, $\lambda $ and  $d_{\varphi}$ represent the wavelength and distance between the satellite and the user $\varphi$, respectively. ${{G_t}}$ is the transmit antenna gain from the satellite antenna to user $\varphi$. ${{G_r}}$ denotes the receive antenna gain of user $\varphi$. In the BH system, the channel gain is dynamically changed according to the time slots, but the channel gain is assumed not to change in time slot $t$.

The active beams are orthogonal to each other and have independent decoding sequences, and thus multiple carriers are assigned by the users within the beam to isolate interference. For the case, e.g., ${\left\| {{h_n}} \right\|^2} > {\left\| {{h_m}} \right\|^2}$, the user $n$ decodes and removes the user $m$'s signal using SIC in the beam. The user $m$ also obtains more power to enhance its performance by the NOMA protocol. Based on previous assumptions, the signal-plus-interference-to-noise ratio (SINR) of the user \emph{m}'s signal ${{x_m}}$ detected by  user \emph{n} is given by
\begin{align}\label{2}
{\gamma _{n \to m}} = \frac{{{P_s}\left\| {{\mathop{\rm diag}\nolimits} \left( {{{\bf{h}}_{n,t}}} \right){{\bf{g}}_{m,t}}} \right\|_2^2{a_{m,t}}}}{{{P_s}\left\| {{\mathop{\rm diag}\nolimits} \left( {{{\bf{h}}_{n,t}}} \right){{\bf{g}}_{n,t}}} \right\|_2^2{a_{n,t}} + {{\bf{I}}_t} + {\omega ^*}{P_s} + {\sigma ^2}}}, \tag{2}
\end{align}
where ${{\bf{I}}_t} = \left\| {{{\bf{I}}_{inter,t}}} \right\|_2^2 + \left\| {{{\bf{I}}_{intra,t}}} \right\|_2^2$, ${{\bf{I}}_{inter,t}} = \sum\limits_{b \in {\cal B}'} {\sum\limits_{\varphi ' \in {\cal M}{\rm{,}}{\cal N}} {diag\left( {{{\bf{h}}_{\varphi ',t}}} \right)} } {{\bf{g}}_{\varphi ',t}}{x_{\varphi '}}{P_s}$ and ${{\bf{I}}_{intra,t}} = \sum\limits_{\varphi ' \in {\cal M}{\rm{,}}{\cal N}\backslash \left\{ \varphi  \right\}} {diag\left( {{{\bf{h}}_{\varphi ',t}}} \right)} {{\bf{g}}_{\varphi ',t}}{x_{\varphi '}}{P_s}$ denote the interference of inter-beams and intra-beam, respectively.

After applying SIC technology, the SINR that user \emph{n} needs to decode itself information is given by
\begin{align}\label{3}
{\gamma _n} = \frac{{{P_s}\left\| {{\mathop{\rm diag}\nolimits} \left( {{{\bf{h}}_{n,t}}} \right){{\bf{g}}_{n,t}}} \right\|_2^2{a_{n,t}}}}{{{{\bf{I}}_t} + {\omega ^*}{P_s} + {\sigma ^2}}}.\tag{3}
\end{align}

The SINR of user \emph{m} to decode the information of itself for the target beam can be expressed as
\begin{align}\label{4}
{\gamma _m} = \frac{{{P_s}\left\| {{\mathop{\rm diag}\nolimits} \left( {{{\bf{h}}_{m,t}}} \right){{\bf{g}}_{m,t}}} \right\|_2^2{a_{m,t}}}}{{{P_s}\left\| {{\mathop{\rm diag}\nolimits} \left( {{{\bf{h}}_{m,t}}} \right){{\bf{g}}_{n,t}}} \right\|_2^2{a_{n,t}} + {{\bf{I}}_t} + {\omega ^*}{P_s} + {\sigma ^2}}}.\tag{4}
\end{align}

We assume that the interference between active beams  hardly  influence the distribution of subcarriers and the order of decoding signals by users. The achievable rate of user $\varphi $  at time slot $t$ for U-NOMA-BH systems can be expressed as
\begin{align}\label{5}
{R_{\varphi t}} = \sum\limits_{\varphi  \in (m,n)} {\sum\limits_{k \in {\cal K}} {\frac{{W{\beta _{\varphi k}}}}{N}{{\log }_2}\left( {1 + {\gamma _{\varphi t}}} \right)} } ,  \tag{5}
\end{align}
where \emph{W} represents the bandwidth per carrier (multi-carrier per beam), \emph{N} is the total number of subcarriers.
$\beta_{\varphi k}$ is defined as:
 \begin{equation}\label{eq}
\beta_{\varphi k}=\left\{\begin{array}{l}1, \text { user } \varphi \text { is assigned at subcarrier } k, \\ 0, \text { otherwise;}\end{array}\right.\nonumber
\end{equation}
 indicating whether subcarrier \emph{k} used by user $\varphi $ or not. The total reachable capacity of NOMA user $\varphi $ is given by
\begin{align}\label{6}
{R_\varphi } = \sum\limits_{t \in {\cal T}} {{R_{\varphi t}}}.\tag{6}
\end{align}

It is worth pointing out that the U-NOMA-BH systems are divided into two settings CD-NOMA-BH ($K \neq 1$) and PD-NOMA-BH ($K=1$) due to the variation in the number of subcarriers. For the particular case PD-NOMA-BH, the optimization of subcarriers will not be considered and equation (4) degenerates to a single-carrier problem.

\section{ PROBLEM FORMULATION}

\subsection{Problem Formulation}
In order to make the actual user satisfaction as high as possible, we formulate a joint optimization problem to compare above the user reachable capacity and user demand calculated, and minimize the gap. The variable of time slot is defined as:
\begin{equation}\label{eq16}
\hspace{0.5cm}\delta_{b t}=\left\{\begin{array}{l}1, \text { beam } b \text { is actived at timeslot } t, \\ 0, \text { otherwise. }\end{array}\right.\nonumber
\end{equation}
The optimization objective function is solved by jointly optimizing time slot $t$, power factor ${a_\varphi }$ and subcarrier variable ${{\beta _{\varphi k}}}$.
Let ${{D_\varphi }}$ represents the request traffic demand of user $\varphi $ within the time slots window. Therefore, we use the well-known mathematical model  $\left(R_{\varphi}-D_{\varphi}\right)^{2}$ to average the contradictions of users traffic mismatch. The optimization problem is formulated as
\begin{align}\label{7}
&{{\cal P}_0}:\mathop {\min }\limits_{{\delta _{bt}},{\beta _{\varphi k}},{a_{\varphi t}}} \sum\limits_{\varphi  \in {\rm{(}}{\cal M}{\rm{,}}{\cal N}{\rm{)}}} {{{\left( {{R_\varphi } - {D_\varphi }} \right)}^2}}\tag{7a}\\
&{\rm{s}}{\rm{.t}}{\rm{.    }}
\quad\sum\limits_{b \in {\cal B}} {{\delta _{bt}}}  \le {B_0},\forall t \in {\cal T}, \tag{7b}\\
&\qquad
 {R_\varphi } \ge R_\varphi ^{\min },\forall \varphi  \in {\cal M},{\cal N},  \tag{7c}\\
&\quad\sum\limits_{k \in {\cal K}} {{\beta _{\varphi k}}}  \le Q{\delta _{bt}},\forall t \in {\cal T},\forall b \in {\cal B},\tag{7d}\\
&\qquad
{\left\| {diag\left( {{{\bf{g}}_{m,t}}{\bf{g}}_{n,t}^{\rm{T}}} \right)} \right\|_0} \ge {\delta _{bt}},\forall \varphi  \in {\cal M},{\cal N},\forall t \in {\cal T} ,  \tag{7e}\\
&\qquad{a_{nt}} \le {a_{mt}},\forall \varphi  \in {\cal M},{\cal N},\forall t \in {\cal T},\tag{7f}
\end{align}
where (7b) confines that the total number of active beams not exceeding ${B_0}$ in any time slot. In (7c), the rate of each user should satisfy the QoS minimum rate requirement in order to maintain a certain level of fairness among users.
 Constraints (7d) restricts the subcarriers length of user ${\varphi }$ at any beam of each time slot to be less than or equal to $Q$, where inactive beams are not scheduled subcarriers. $Q$ is defined as the maximum value of the user $\varphi $ assigned to the subcarriers in the sparse matrix. To ensure the transmission of paired NOMA users on the same subcarrier, (7e) is the carrier vector transpose of user $m$ multiplied by user $n$ and then the diagonal element of the result is obtained. By exploiting the vector zero-norm\cite{Li2021}, the number of intersecting non-zeros for users $m$ and $n$ is made no less than ${\delta _{bt}}$ .
Considering the fairness of the users in (7f), the power allocation coefficient of the user \emph{n} should be less than the user \emph{m} in each time slot, and the sum of the power factors of the users in the paired NOMA user is equal to one. In order to avoid the waste of resources, we usually make the user ${\varphi}$ achievable  capacity not exceed the capacity requirement, i.e., $R_\varphi  < {D_\varphi }$.

Further analyses show that the optimization objectives in $\mathcal{P}_{0}$ having mutual constraints. When the beam is activated, the power allocation of NOMA needs to consider the matching between users, and the subcarriers allocation naturally requires user pairs to transmit signals on the same subcarrier in a single beam, which leads to complex design problems. In addition, it is the problem of power allocation among NOMA users, where different power allocations will affect the sum of intra-beam interference. To improve the optimization result, the change of mutual interference between users is recognized at the same time slot when subcarriers allocation selects the subcarrier-channel combination of the largest user.
Formulating a high-quality BH system can effectively improve the optimal results, where the design of the power and subcarriers in the multi-dimension is particularly important for the BH system.
When the joint optimization of the three variables is directly performed, more complex dimensionality problems will arise. Hence the necessary intra-beam and inter-beams problems are analyzed to realize an effective U-NOMA-BH systems separately. The spatial properties of variables are used for step-by-step design, which simplifies the design difficulty of the algorithm.
\subsection{ Subcarrier and Power Optimization in Solving $\mathcal{P}_{0}$}
The begin of designing the optimization scheme, we need to solve the non-convexity of $\mathcal{P}_{0}$. Since the objective function and the constraints both have nonlinear functions,  $\mathcal{P}_{0}$ is a nonlinear programming problem. Deciding nonlinear programming directly is very difficult, so we relax the optimization problem to solve the MINCP problem. $\mathcal{P}_{0}$ has discrete variables in (7c) and (7d), and still needs to address its non-convexity. We relax the variables  $\delta_{b t}$ into continuous variables to reduce the complexity of the solution.

In the power allocation algorithm, fixing subcarriers allocation is assumed across all users. In general, decoupling the numerator and denominator are applying to solve non-convex problems, and Dinkelbach's transform is still very competitive in terms of power allocation\cite{Zappone2017}. Compared with other solutions, the Dinkelbach's transform does not divide the result, but it gives and moves an answer according to a better solution to approach the optimal way, which is  more sufficient for each judgment. Now turning our attention to problem (7) again, the target is the optimization of $a_{\varphi t}$ by fixing $\theta_{\varphi}$. In particular, $R_{\varphi}$ changed from fractional format to the following
\begin{align}\label{8}
{\cal L}_{\varphi }^R\left( {{\theta _\varphi },{a_{\varphi t}}} \right) &= \log \{ 1 + \left\| {{\mathop{\rm diag}\nolimits} \left( {{{\bf{h}}_{n,t}}} \right){{\bf{g}}_{n,t}}} \nonumber \right\|_2^2{a_n}\zeta \\\nonumber
& + \left\| {{\mathop{\rm diag}\nolimits} \left( {{{\bf{h}}_{m,t}}} \right){{\bf{g}}_{m,t}}} \right\|_2^2{a_m}\zeta \\\nonumber
& + \left\| {{\mathop{\rm diag}\nolimits} \left( {{{\bf{h}}_{n,t}}} \right){{\bf{g}}_{n,t}}} \right\|_2^2\left\| {{\mathop{\rm diag}\nolimits} \left( {{{\bf{h}}_{m,t}}} \right){{\bf{g}}_{n,t}}} \right\|_2^2a_n^2  \\\nonumber
&+ \left\| {{\mathop{\rm diag}\nolimits} \left( {{{\bf{h}}_{n,t}}} \right){{\bf{g}}_{n,t}}} \right\|_2^2\left\| {{\mathop{\rm diag}\nolimits} \left( {{{\bf{h}}_{m,t}}} \right){{\bf{g}}_{m,t}}} \right\|_2^2{a_m}{a_n} \\\nonumber
&- {\theta _\varphi }[\zeta (\zeta  + \left\| {{\mathop{\rm diag}\nolimits} \left( {{{\bf{h}}_{m,t}}} \right){{\bf{g}}_{n,t}}} \right\|_2^2{a_n} + \zeta )]\} ,\tag{8}
\end{align}
where $\zeta  = {{\bf{I}}_t} + {\omega ^*}P + {\sigma ^2}$, ${\theta _\varphi } > 0$. Due to the non-convexity of $\mathcal{P}_{0}$, ${\cal L}_\varphi ^R$ is replaced by ${R_{\varphi }}$ to transform the optimization problem. With fixed $\theta_{\varphi}$, ${\cal L}_\varphi ^R$ is a concave function. In each iteration, the procedures of alternatively update $\theta_{\varphi}$  and ${a_{\varphi t}}$ are discussed, respectively, where $\theta_{\varphi}$ is updated with optimized power allocation factor. With fixed ${\cal L}_\varphi ^R$, the optimal $\theta_{\varphi}$ is derived by (10). Thus, the optimization problem can be formulated as\cite{Wang2012SCA}
\begin{align}\label{9}
&{{\cal P}_1}:{\min _{{a_{\varphi t}}}}\sum\limits_{\varphi  \in ({\cal M},{\cal N})} {{{\left( {{\cal L}_\varphi ^R - {D_\varphi }} \right)}^2}}  \tag{9a}\\
&{\rm{s}}{\rm{.t}}{\rm{.    }}\qquad(7e)\;,\;(7f).\tag{9b}
\end{align}

As a further advance, we assume that the users pair power within a beam is uniformly distributed. The subcarriers allocation is optimized within each beam according to (7e) and (7d). Firstly, subcarriers allocation for CD-NOMA-BH users is a matching problem of user pairs. The combined user pairs sharing the same subcarriers have intra-beam interference with other users due to the sparse matrix. Consequently, we pair fully orthogonal users in $\mathbf{G}_{K \times M}$ with other users, which reduces code complexity and intra-beam interference. Specifically, if $MQ \le K$ , picking ${{\bf{g}}_{\varphi ,t}}$ that are orthogonal to each other is indicated to $v$. Matching user pairs according to $v$, selecting ${{\bf{g}}_{\varphi ',t}}$ and ${{\bf{g}}_{\varphi ,t}}$ to have mutually non-orthogonal subcarriers is expressed as
\begin{align}\label{11}
\left[ {\begin{array}{*{20}{c}}
{{\beta _{n1}}}\\
{{\beta _{nk}}}\\
 \vdots \\
0
\end{array}} \right]{\left[ {\begin{array}{*{20}{c}}
0\\
{{\beta _{mk}}}\\
 \vdots \\
{{\beta _{mK}}}
\end{array}} \right]^{\rm{T}}} \Rightarrow {\left[ {\begin{array}{*{20}{c}}
0&0& \cdots &1\\
1&{{\beta _{\varphi k}}}& \cdots &0\\
 \vdots & \vdots & \ddots & \vdots \\
0&0& \cdots &{{\beta _{\varphi 'K}}}
\end{array}} \right]_{K \times M}}.\tag{11}
\end{align}
If the users are not satisfied (12), a matching exchange is performed. Finally, the matched the set of user ${v}$ and ${v'}$ are combined $\mathbf{G}_{K \times M}$.

\begin{figure*}[!t]
\normalsize
\begin{align}\label{10}
{\theta _\varphi } = \frac{{\left[ {\left\| {{\mathop{\rm diag}\nolimits} \left( {{{\bf{h}}_{m,t}}} \right){{\bf{g}}_{n,t}}} \right\|_2^2{a_n} + \zeta } \right]\left\| {{\mathop{\rm diag}\nolimits} \left( {{{\bf{h}}_{n,t}}} \right){{\bf{g}}_{n,t}}} \right\|_2^2{a_n} + \left[ {{a_m}\zeta  + \left\| {{\mathop{\rm diag}\nolimits} \left( {{{\bf{h}}_{n,t}}} \right){{\bf{g}}_{n,t}}} \right\|_2^2{a_m}{a_n}} \right]\left\| {{\mathop{\rm diag}\nolimits} \left( {{{\bf{h}}_{m,t}}} \right){{\bf{g}}_{m,t}}} \right\|_2^2}}{{\zeta \left( {\zeta  + \left\| {{\mathop{\rm diag}\nolimits} \left( {{{\bf{h}}_{m,t}}} \right){{\bf{g}}_{n,t}}} \right\|_2^2{a_n} + \zeta } \right)}}.\tag{10}
\end{align}
\hrulefill \vspace*{0pt}
\end{figure*}

In the Algorithm 1, according to (8) and (11), a joint optimization scheme of power and subcarriers is designed through fixed time slots. In the subcarriers optimization process, a non-orthogonal user groups are defined, and  an orthogonal user that is expected to be matched is searched, which reduces the number of iterations for direct user pairing. Compared with exhaustive search, the complexity is reduced by $\left[ {M - (K/Q)} \right]!/M!$ times. In the optimal allocation of the power factor, the power is optimized using the subcarrier and user pair clusters obtained above.
Based on (8) and (10) we derive a better solution with the initial value of the power factor. The power allocation algorithm can find the global optimal solution, and the need to save the solution has little effect on the time complexity. The complexity of Algorithm 1 mainly from the circular iterative process and the optimization of power. To optimize  ${a_{\varphi t}}$, we adopt the interior point method to solve the complex $\mathcal{P}_1$. The complexity of the interior point method is defined as ${\cal O}\left( {\psi \log \left( {\frac{1}{\varepsilon }} \right)} \right)$,  where $\psi  > 0$ represents the parameter for self-concordant barrier and $\varepsilon  > 0$ is the precision\cite{2010inter}. In conclusion, the user pairing carrier-based cycle is $B \times (B - 1)$ times. Thus, the complexity of Algorithm 1 is ${\cal O}\left( {{N_1}\left( {B \times \left( {B - 1} \right) + {N_2}\psi \left( {\log \left( {\frac{1}{\varepsilon }} \right)} \right)} \right)} \right)$, where ${N_1}$ and ${N_2}$ are the maximum number of iterations.

\begin{algorithm}\label{alg:EBJT}
\caption{Power and Subcarrier joint optimization scheme}
\hspace*{0.001in} {\bf Input:}\;flexible $\theta_{\varphi}$
\begin{algorithmic}
\STATE  Initialize: time slot, power factor\\
\hspace*{0.01in}  {\bf repeat} \\
\hspace*{0.02in}  {\bf for} $i = 1,2....B$\\
\hspace*{0.2in}   {\bf First round:}  Perform user pairing and select complementary orthogonal user sets $v$\\
\hspace*{0.2in}   {\bf Second round:}  Find orthogonal users between users according to $v$\\
\hspace*{0.3in}   {\bf for} $j=1$:combinations of remaining users\\
\hspace*{0.35in}   {\bf if}  ${\left\| {{{\bf{g}}_{m,t}}{\bf{g}}_{n,t}^{\rm{T}}} \right\|_0} \le 1$\\
\hspace*{0.45in}   break\\
\hspace*{0.35in}   {\bf else} \\
\hspace*{0.45in}   Identifies as a NOMA user pair\\
\hspace*{0.35in}   {\bf end if}\\
\hspace*{0.3in}    {\bf end for}\\
\hspace*{0.2in}    Calculate (7a) to get the user allocation and ${{\beta _{\varphi k}}}$\\
\hspace*{0.02in}  {\bf end for} \\
\hspace*{0.02in}   Consider ${{\beta _{\varphi k}}}$ as the initial value of subcarrier allocation in optimizing power.\\
\hspace*{0.02in}    Update $\theta_{\varphi}$ by (10)\\
\hspace*{0.02in} {\bf repeat}\\
\hspace*{0.03in}  Optimize $a_{\varphi t}$ and $\theta_{\varphi}$ by solving $\mathcal{P}_1$\\
\hspace*{0.03in}  $a_{\varphi t}^* = \arg \min \left[ {\mathop {\min }\limits_{{a_{\varphi t}}} \sum\limits_{\varphi  \in ({\cal M},{\cal N})} {{{\left( {{\cal L}_\varphi ^R - {D_\varphi }} \right)}^2}} } \right]$\\
\hspace*{0.03in}   ${{\theta '}_\varphi } = {{\theta '}_\varphi }\left( {a_{\varphi t}^*} \right)$ \\
\hspace*{0.02in} {\bf until convergence} \\
\hspace*{0.03in} Obtain ${{a_{\varphi t}}}$, and replace the initial value of the power factor\\
\hspace*{0.01in} {\bf until convergence} \\
\STATE Obtain $a_{\varphi t}$ and $\beta_{\varphi k}$  for a single timeslot\\
\end{algorithmic}
\hspace*{0.001in} {\bf Output:}\;$R_{\varphi}$, $a_{\varphi t}$\\
\end{algorithm}

It is worth noting that the optimization of PD-NOMA-BH is still special. Since the single-carrier transmission with $K=1$, we cannot utilize subcarrier intersection for user pairing. The set of users is divided according to the order of channel gain, which can be changed from an exhaustive search of all users to a match of ${\cal M}$ and ${\cal N}$. Inter-users interference within a beam is also increased due to a single carrier. Hence the optimization scheme is more sensitive to interference between users. Then the PD-NOMA-BH algorithm abandons (7d), (7e) and only optimizes the power allocation. The problem is formulated as:

\begin{align}\label{12}
&{{\cal P}_2}:{\min _{{a_{\varphi t}}}}\sum\limits_{\varphi  \in ({\cal M},{\cal N})} {{{\left( {{\cal L}_\varphi ^R - {D_\varphi }} \right)}^2}} \tag{12a}\\
&{\rm{s}}{\rm{.t}}{\rm{.    }}
\quad (7d), (7e),\tag{12b}\\
&\qquad{a_{nt}} \le {a_{mt}},\forall \varphi  \in {\cal M},{\cal N},\left( {{\cal M},{\cal N}} \right) \in \Omega ,\forall t \in {\cal T}.\tag{12c}
\end{align}
\subsection{ Optimization of time slots allocation}
In this subsection, we apply iterations of users traffic demands to solve the BH decision problem. The allocation of time slots is related to the interference between beams. Assuming that the user channel gain does not change in each time slot, the interference between beams will cause the SINR to deteriorate. Given the power factor and subcarrier assignments derived above, we relax the variable $\delta_{b t}$ to optimize the time slots assignment. The design of BH turns the optimization problem into a sub-problem within each time slot. The remaining demand for user $\varphi$ in time period \emph{t} is given by
\begin{align}\label{13}
{Y_\varphi } = {D_\varphi } - \sum\limits_{\tau  = 0}^{t - 1} {{R_{\varphi \tau }}}  .\tag{13}
\end{align}
Since ${{\cal P}_0}$ is MINCP and difficult to calculate, we further relax $\delta_{b t}$ into continuous variables, i.e., $0 \le {\delta _{\varphi t}} \le 1$. The resource allocation problem for the current $t$ time slot is expressed as
\begin{align}\label{14}
&{{\cal P}_3}:\mathop {\min }\limits_{{\delta _{bt}},{\beta _{\varphi k}},{a_{\varphi t}}} \sum\limits_{\varphi  \in ({\cal M},{\cal N})} {{{\left( {{\cal L}_\varphi ^R - {Y_\varphi }} \right)}^2}}   \tag{14a}\\
&{\rm{s}}{\rm{.t}}{\rm{.    }}\quad \sum_{b \in \mathcal{B}} \delta_{b t} \leq B_{0}, \forall t \in \mathcal{T},\tag{14b}\\
&\qquad \sum_{t \in \mathcal{T}} \delta_{b t} \leq T, \forall b \in \mathcal{B},\tag{14c}\\
&\qquad\sum_{\varphi \in(\mathcal{M}, \mathcal{N})} \beta_{\varphi k} \leq Q \delta_{b t}, \forall t \in \mathcal{T}, \forall b \in \mathcal{B},\tag{14d}\\
&\qquad\left\|\mathbf{g}_{m, t} \mathbf{g}_{n, t}^{\mathrm{T}}\right\|_0 \geq \delta_{b t} \forall \varphi \in \mathcal{M}, \mathcal{N}, \forall t \in \mathcal{T},\tag{14e}\\
&\qquad a_{nt} \leq a_{mt}, \forall \varphi \in \mathcal{M}, \mathcal{N}, \forall t \in \mathcal{T}.\tag{14f}
\end{align}

\begin{algorithm}\label{time}
\caption{Timeslots optimization scheme}
\hspace*{0.001in} {\bf Input:}\;$a_{\varphi t}$, $\beta_{\varphi k}$
\begin{algorithmic}
\STATE  Optimize $a_{\varphi t}$ by solving $\mathcal{P}_{3}$ via Algorithm 1\\
\hspace*{0.02in} {\bf repeat}\\
\hspace*{0.03in} {\bf for}\;$t=1, \ldots \ldots, T$\\
\hspace*{0.08in} Calculate the time slot required by the user $\varphi $ according to the traffic demand ${Y_\varphi }$, and mark the beam where it is located as $j$\\
\hspace*{0.1in} {\bf for}\;$b=1, \ldots \ldots, B$\\
\hspace*{0.2in} {\bf if}  ${\delta _{bt}}(j) > 1$\\
\hspace*{0.3in} ${\delta _{bt}}(j) = {\delta _{bt}}(j) - 1$\\
\hspace*{0.2in} {\bf else}\\
\hspace*{0.2in} The working beam sets will remove the beam $j$ with time slot less than 1 and seek the next beam\\
\hspace*{0.3in} ${Y_{\varphi j}} =  + \infty $\\
\hspace*{0.2in} {\bf end if}\\
\hspace*{0.2in} Select CD-NOMA-BH user pairs via Algorithm 1\\
\hspace*{0.1in} {\bf end for}\\
\hspace*{0.08in} Optimize time slot allocation $\delta_{b t}$ is solved by solving $\mathcal{P}_3$\\
\hspace*{0.03in} {\bf end for}\\
\hspace*{0.02in} Update ${Y_\varphi }$ by (13)\\
\hspace*{0.02in} {\bf until convergence}\\
\STATE Obtain $\delta_{b t}$ and $\beta_{\varphi k}$ by solving $\mathcal{P}_{3}$ with the determined time slot assignment\\
\end{algorithmic}
\hspace*{0.001in} {\bf Output:}\;${a_{\varphi t}}$, $\delta_{b t}$, $\beta_{\varphi k}$\\
\end{algorithm}

The algorithm of BH pattern consists of two steps: the pre-allocation and dynamic allocation of time slots. In the initial phase, per beam is assigned a time slot to ensure that time slots are not allocated to user beams that are rarely requested. After pre-allocation, the dynamic allocation applies continuous variable $\delta_{b t}$ to find the set of beams that requires the least number of time slots, and allocates the current time slot to the set of beams. The satellites serve users with the smallest capacity optimized by Algorithm 1 in each iteration. Specifically, Algorithm 2 works by updating per user request traffic and continuing the search in the next time slot. By calculating the total number of time slots to reach the minimum range, it amplifies user demand so that it is no longer falsely illuminated. The algorithm terminates and is not updated when the number of allocated time slots reaches the time slots window. The method ensures that satisfiable beams are illuminated within the maximum window, maximizing user satisfaction and greatly improving the system time slot utilization.

In general, the algorithm 2 is applied to algorithm 1 $B \times T$ times, and it has complexity denoted as ${\cal O}\left( {{N_3}\left( {B \times T} \right){N_1}\left( {B \times \left( {B - 1} \right) + {N_2}\psi \left( {\log \left( {\frac{1}{\varepsilon }} \right)} \right)} \right)} \right)$, where ${N_3}$ is the maximum number of time slot optimization iterations. In continuation, the dynamic demand change problem of the above optimization problem is analyzed. To describe the modification of the user iterative demand difference that may occur in the optimization process, the following two situations are required:
\begin{itemize}
\item Since the user demand changes with time slot, the previously obtained traffic demand gap may be changed in the next time slot.
\item Based on discrete variables whose result is 0 or 1 for ${{\delta _{bt}}}$, beam-level time slots assignments lead to differences in the user traffic demands.
\end{itemize}

\section{Performance Analysis}
In this section, we further analyze the NOMA user pairs within the beam exploiting the CD/PD-NOMA-BH algorithm previously described. According to the above algorithm, a pair of users, i.e., user $n$ and user $m$ are selected from BCU and BEU to carried out NOMA protocol, respectively. To deeply explore the reliable rate of user services in the BH system, the outage probabilities of user \emph{n} and user \emph{m} for the target beam are studied based on the aforementioned algorithm.
\subsection{The Outage Probability of User $n$}
The outage for the user \emph{n} of BCU can appear in the following two cases\cite{Yue2022mn}: i) The user \emph{n} can only decode its own signal in NOMA-BH user pair; and ii) The user \emph{n} successfully decodes user \emph{m}'s signal, but fails to decode its own signal after SIC.

To ensure high quality of service, we introduce the parameter $\kappa $ to explain the number of time slots shared by the target beam and the interference beam in the BH, where the data rate of user within $\kappa $ must not be less than the set target rate.  Assuming that target SINR threshold ${\varepsilon _\varphi } = {2^{\frac{{{R_\varphi }}}{W}}} - 1$, where $R_{\varphi}$ is the achievable target rate. An outage event is defined as the received SINR falls below the target SINR threshold. As a result, the outage probability of user \emph{n} in CD/PD-NOMA-BH systems can be expressed as
 \begin{align}\label{15}
 P_n^{BH} = 1 - \Pr \left( {{\gamma _{n \to m}} > {\varepsilon _m},{\gamma _n} > {\varepsilon _n}} \right).\tag{15}
\end{align}

\begin{theorem}\label{Theorem:CD-NOMA:the COP of near user for Case1 with ipSIC}
The closed-form expression for the outage probability of the investigated user n with ipCSI can be expressed as in CD-NOMA-BH system
\begin{align}\label{16}
P_{n,ipCSI}^{CD - BH} = \sum\limits_{b' \in {\cal B}\backslash \left\{ b \right\}} {{e^{ - \frac{{{\Lambda _1}}}{\chi }}}} \sum\limits_{i = 1}^{K - 1} {\frac{1}{{i!}}{{\left( {\frac{{{\Lambda _1}}}{\chi }} \right)}^i}{P_{avg}}{{\left( {b'} \right)}_\kappa }}   ,\tag{16}
\end{align}
where $\chi=\sqrt{G_{r}} \sqrt{\left(\frac{\lambda}{4 \pi d_{\mathrm{m}}}\right)^{2}} \sqrt{G_{t}}$, ${\Omega _1} = \frac{{{\varepsilon _m}{\omega ^*}}}{{{a_{mt}} - {\varepsilon _m}{a_{nt}}}}$, ${\Omega _2} = \frac{{{\varepsilon _n}{\omega ^*}}}{{{a_{nt}}}}$, ${\tau _1} = \frac{{{\varepsilon _m}}}{{\rho \left( {{a_{mt}} - {\varepsilon _m}{a_{nt}}} \right)}}$, ${\tau _2} = \frac{{{\varepsilon _n}}}{{\rho {a_{nt}}}}$, ${\Lambda _1} \buildrel \Delta \over = \max \left( {{\tau _1} + {\Omega _1},{\tau _2} + {\Omega _2}} \right)$, $\rho  = {P_s}/{\sigma ^2}$. ${{P_{avg}}{{\left( {b'} \right)}_\kappa }}$  is the average probability of ${b'}$ inter-beams interference in the direction of the target beam for $\kappa $.
\begin{proof}\label{Proof:CD-NOMA:the COP of near user for Case1 with ipSIC}
 Considering a two-user in one beam case, the user m and user n are paired together to perform NOMA protocol. Supposing $\left\|\operatorname{diag}\left(\mathbf{h}_{n, t}\right) \mathbf{g}_{m, t}\right\|_{2}^{2}$ and $\left\|\operatorname{diag}\left(\mathbf{h}_{m, t}\right) \mathbf{g}_{n, t}\right\|_{2}^{2}$ follow the same distribution for the satellite links, where $\left\|\operatorname{diag}\left(\mathbf{h}_{m, t}\right) \mathbf{g}_{m, t}\right\|_{2}^{2}=\sqrt{G_{r}} \sqrt{\left(\frac{\lambda}{4 \pi d_{m}}\right)^{2}} \sqrt{G_{t}} \sum_{k=1}^{K}\left|g_{m k} \bar{h}_{m k}\right|^{2}$. It is observed that $Z=\sum_{k=1}^{K}\left|g_{m k} \bar{h}_{m k}\right|^{2}$ is subject to a Gamma distribution with the parameters of $(K, 1)$. The corresponding CDF and PDF of $Z$ are given by
$F_{Z}(z)=1-e^{-z} \sum_{i=0}^{K-1} \frac{z^{i}}{i !}$
and
$f_{Z}(z)=\frac{z^{K-1} e^{-z}}{(K-1) !}$, respectively \cite{Liu2017LS}. The probability of co-channel interference has derived to provide a expression by \cite{Howitt1998}
\begin{align}\label{19}
P_{a v g}\left(b^{\prime}\right)_{\kappa}=\frac{1}{\pi r^{2}} \int_{0}^{2 \pi} \int_{0}^{r} P\left(b^{\prime} \mid \vartheta, r\right)_{\kappa} d \vartheta d r,\tag{17}
\end{align}
where $\vartheta $ is the pointing direction of the target beam, $r$ is the radius of the target beam. ${{P_{avg}}{{\left( {b'} \right)}_\kappa }}$ is evaluated using a Monte Carlo simulation. The value of ${b'}$ is determined for each iteration of the Monte Carlo simulation for $\kappa$.

By definition, (15) denotes the complementary event at user n and is calculated as
\begin{align}\label{20}
P_{n,ipCSI}^{CD - BH} \nonumber
 = &1 - \Pr \left( {{\gamma _{n \to m}} > {\varepsilon _m},{\gamma _n} > {\varepsilon _n}} \right)\\\nonumber
 = &1 - \sum\limits_{b' \in {\cal B}\backslash \left\{ b \right\}} {\Pr \left( {{{\left| {{h_n}} \right|}^2} > {\Lambda _1}} \right)}  \\\nonumber
 &\times \frac{1}{{\pi {r^2}}}\int_0^{2\pi } {\int_0^r {P{{\left( {b'|\vartheta ,r} \right)}_\kappa }} } d\vartheta dr\\
 = &1 - \sum\limits_{b' \in {\cal B}\backslash \left\{ b \right\}} {{P_{avg}}{{\left( {b'} \right)}_\kappa }\int_{{\Lambda _1}}^\infty  {{f_{{{\left| {{h_n}} \right|}^2}}}(y|b')dy} }.\tag{18}
\end{align}
Substituting (3) into (17) and the PDF of $Z$ can be obtained and the proof is completed.
\end{proof}
\end{theorem}
\begin{corollary}\label{Corollary:n-pCSI-CD-NOMA-BH}
For the special case of ${\omega ^*} = 0$, the closed-form expression of outage probability for user n with pCSI in CD-NOMA-BH system can be given by
\begin{align}\label{21}
P_{n,pCSI}^{CD - BH} = \sum\limits_{b' \in {\cal B}\backslash \left\{ b \right\}} {{e^{ - \frac{{{\Lambda _2}}}{\chi }}}} \sum\limits_{i = 1}^{K - 1} {\frac{1}{{i!}}{{\left( {\frac{{{\Lambda _2}}}{\chi }} \right)}^i}{P_{avg}}{{\left( {b'} \right)}_\kappa }} ,\tag{19}
\end{align}
where ${\Lambda _2} \buildrel \Delta \over = \max \left( {{\tau _1},{\tau _2}} \right)$.
\end{corollary}

\begin{corollary}\label{Corollary:PD-NOMA:the PD of near user n for case1}
For the special  case of $K=1$, the closed-form expression of outage probability for user n with ipCSI/pCSI in PD-NOMA-BH system can be represented as
\begin{align}\label{23}
P_{n,ipCSI}^{PD - BH} = \sum\limits_{b' \in {\cal B}\backslash \left\{ b \right\}} {{e^{ - \frac{{{\Lambda _1}}}{\chi }}}} {P_{avg}}{\left( {b'} \right)_\kappa },\tag{20}
\end{align}
and
\begin{align}\label{24}
P_{n,pCSI}^{PD - BH} = \sum\limits_{b' \in {\cal B}\backslash \left\{ b \right\}} {{e^{ - \frac{{{\Lambda _2}}}{\chi }}}} {P_{avg}}{\left( {b'} \right)_\kappa },\tag{21}
\end{align}
respectively.
\end{corollary}
\subsection{The Outage Probability of User $m$}
According to NOMA protocol, the outage event occurs for user \emph{m} when it is unable to decode its own signal at both time slots. Thus, the outage probability of user \emph{m} in CD/PD-NOMA-BH systems can be expressed as
\begin{align}\label{25}
P_m^{BH} = 1 - \Pr \left( {{\gamma _m} > {\varepsilon _m}} \right).\tag{22}
\end{align}

\begin{theorem}\label{Theorem:CD-NOMA:the COP of near user for Case1 with ipSIC}
The closed-form expression for the outage probability of the investigated user m with ipCSI can be expressed as in CD-NOMA-BH system
\begin{align}\label{26}
P_{m,ipCSI}^{CD - BH} \nonumber
=& \sum\limits_{b' \in {\cal B}\backslash \left\{ b \right\}} {{P_{avg}}{{\left( {b'} \right)}_\kappa }{e^{ - \frac{{{\omega ^*}{\varepsilon _m} + {\tau _m}\left( {{a_{mt}} - {\varepsilon _m}{a_{nt}}} \right)}}{{\chi \left( {{a_{mt}} - {\varepsilon _m}{a_{nt}}} \right)}}}}}  \\\nonumber
&\times \sum\limits_{i = 1}^{K - 1} {\frac{1}{{i!}}{{\left[ {\frac{{{\omega ^*}{\varepsilon _m} + {\tau _m}\left( {{a_{mt}} - {\varepsilon _m}{a_{nt}}} \right)}}{{\chi \left( {{a_{mt}} - {\varepsilon _m}{a_{nt}}} \right)}}} \right]}^i}} ,\tag{23}
\end{align}
where ${\tau _m} = \frac{{{\varepsilon _m}}}{{\rho \left( {{a_{mt}} - {\varepsilon _m}{a_{nt}}} \right)}}$.
\begin{proof}\label{Proof:CD-NOMA:the COP of near user for Case1 with ipSIC}
By definition, (23) denotes outage events. The process calculated is given by
\begin{align}\label{27}
P_{m,ipCSI}^{CD - BH}  \nonumber
=& 1 - \Pr \left( {{\gamma _m} > {\varepsilon _m}} \right)\\\nonumber
=& 1 - \sum\limits_{b' \in {\cal B}\backslash \left\{ b \right\}} {\Pr \left( {{{\left| {{h_n}} \right|}^2} > {\tau _m}} \right)}  \\\nonumber
& \times \frac{1}{{\pi {r^2}}}\int_0^{2\pi } {\int_0^r {P{{\left( {b'|\vartheta ,r} \right)}_\kappa }} } d\vartheta dr \\\nonumber
 = &1 - \sum\limits_{b' \in {\cal B}\backslash \left\{ b \right\}} {{P_{avg}}{{\left( {b'} \right)}_\kappa }\int_{{\tau _m}}^\infty  {{f_{{{\left| {{h_m}} \right|}^2}}}} (y|b')dy} .\tag{24}
\end{align}

Combining (4), (22) and the PDF of $Z$ can be obtained and the proof is completed.
\end{proof}
\end{theorem}

\begin{corollary}\label{Corollary:m-pCSI-CD-NOMA-BH}
For the special case of ${\omega ^*} = 0$, the closed-form expression of outage probability for user m with pCSI in CD-NOMA-BH can be written as
\begin{align}\label{28}
P_{m,pCSI}^{CD - BH} \nonumber
=& \sum\limits_{b' \in {\cal B}\backslash \left\{ b \right\}} {{e^{ - \frac{{{\tau _m}\left( {{a_{mt}} - {\varepsilon _m}{a_{nt}}} \right)}}{{\chi \left( {{a_{mt}} - {\varepsilon _m}{a_{nt}}} \right)}}}}}\\\nonumber
 &\times \sum\limits_{i = 1}^{K - 1} {\frac{1}{{i!}}{{\left[ {\frac{{{\tau _m}\left( {{a_{mt}} - {\varepsilon _m}{a_{nt}}} \right)}}{{\chi \left( {{a_{mt}} - {\varepsilon _m}{a_{nt}}} \right)}}} \right]}^i}} {P_{avg}}{\left( {b'} \right)_\kappa }.\tag{25}
\end{align}
\end{corollary}

\begin{corollary}\label{Corollary:PD-NOMA:the PD of near user m for case1}
For the special case of $K=1$, the closed-form expression of outage probability for user m with ipCSI/pCSI in PD-NOMA-BH system can be given by
\begin{align}\label{30}
P_{m,ipCSI}^{PD - BH} = \sum\limits_{b' \in {\cal B}\backslash \left\{ b \right\}} {{e^{ - \frac{{{\omega ^*}{\varepsilon _m} + {\tau _m}\left( {{a_{mt}} - {\varepsilon _m}{a_{nt}}} \right)}}{{\chi \left( {{a_{mt}} - {\varepsilon _m}{a_{nt}}} \right)}}}}{P_{avg}}{{\left( {b'} \right)}_\kappa },}   \tag{26}
\end{align}
and
\begin{align}\label{31}
P_{m,pCSI}^{PD - BH} = \sum\limits_{b' \in {\cal B}\backslash \left\{ b \right\}} {{e^{ - \frac{{{\tau _m}\left( {{a_{mt}} - {\varepsilon _m}{a_{nt}}} \right)}}{{\chi \left( {{a_{mt}} - {\varepsilon _m}{a_{nt}}} \right)}}}}{P_{avg}}{{\left( {b'} \right)}_\kappa },}  \tag{27}
\end{align}
respectively.
\end{corollary}
\subsection{Diversity Order Analyses}
To obtain more sufficient insights, the diversity order is usually a crucial parameter to evaluate the outage performance, which  expresses the slope of the curves for outage probabilities varying as SNR changes. Precisely speaking, the diversity order is defined as
\begin{align}\label{32}
D=-\lim _{\rho \rightarrow \infty} \frac{\log \left(P^{\infty}(\rho)\right)}{\log \rho}, \tag{28}
\end{align}
where $P^{\infty}(\rho)$ denotes the asymptotic outage probability at high SNRs.
\begin{corollary}\label{Corollary:CD-NOMA:the COP of near user for case1}
Based on analytical results in (16), (19) and (28), when $\rho \rightarrow \infty$, the asymptotic outage probability of user n with ipCSI/pCSI at high SNR for CD-NOMA-BH system can be given by
\begin{align}\label{33}
P_{n,ipCSI}^{CD - BH,\infty } = \left[ {1 - \frac{1}{{K!}}{{\left( {\frac{{{\Lambda _1}}}{\chi }} \right)}^K}} \right]\sum\limits_{b' \in {\cal B}\backslash \left\{ b \right\}} {{P_{avg}}{{\left( {b'} \right)}_\kappa }}  ,\tag{29}
\end{align}
and
\begin{align}\label{34}
P_{n,pCSI}^{CD - BH,\infty } = \left[ {1 - \frac{1}{{K!}}{{\left( {\frac{{{\Lambda _2}}}{\chi }} \right)}^K}} \right]\sum\limits_{b' \in {\cal B}\backslash \left\{ b \right\}} {{P_{avg}}{{\left( {b'} \right)}_\kappa }}   ,\tag{30}
\end{align}
respectively.
\end{corollary}
\begin{remark}
Upon substituting (29) and (30) into (28), user n gets zero diversity order as ipCSI applied, which achieves the same results as terrestrial NOMA communications. The diversity orders of the user n with pCSI is K.
\end{remark}
\begin{corollary}\label{Corollary:PD-NOMA:the COP of near user for case1}
For the special case of $K=1$, based on analytical result in (20) and (21), when $\rho \rightarrow \infty$, the asymptotic outage probability of the user $m$ with ipCSI/pCSI for PD-NOMA-BH system can be given by
\begin{align}\label{35}
P_{n,ipCSI}^{PD - BH,\infty } =  - \left( {\frac{{{\Omega _n}}}{\chi } + \frac{{{\tau _n}}}{\chi }} \right)\sum\limits_{b' \in {\cal B}\backslash \left\{ b \right\}} {{P_{avg}}{{\left( {b'} \right)}_\kappa }}  ,\tag{31}
\end{align}
and
\begin{align}\label{36}
P_{n,pCSI}^{PD - BH,\infty } = \left( {1 - \frac{{{\Lambda _2}}}{\chi }} \right)\sum\limits_{b' \in {\cal B}\backslash \left\{ b \right\}} {{P_{avg}}{{\left( {b'} \right)}_\kappa }}  ,\tag{32}
\end{align}
respectively.
\end{corollary}
\begin{remark}
Upon substituting (31) and (32) into (28), the diversity orders of the user n with ipCSI/pCSI are 0 and 1, respectively.
\end{remark}

\begin{corollary}\label{Corollary:PD-NOMA:the COP of near user for case1}
Based on analytical results in (23), (25) and (28), when $\rho \rightarrow \infty$, the asymptotic outage probability of user m with ipCSI/pCSI at high SNR for CD-NOMA-BH system can be given by
\begin{align}\label{37}
P_{m,ipCSI}^{CD - BH,\infty } \nonumber
=& \sum\limits_{b' \in {\cal B}\backslash \left\{ b \right\}} {{P_{avg}}{{\left( {b'} \right)}_\kappa }} \\\nonumber
 &\times \left\{ {1 - \frac{1}{{K!}}{{\left[ {\frac{{{\omega ^*}{\varepsilon _m} + {\tau _m}\left( {{a_{mt}} - {\varepsilon _m}{a_{nt}}} \right)}}{{\chi \left( {{a_{mt}} - {\varepsilon _m}{a_{nt}}} \right)}}} \right]}^K}} \right\},\tag{33}
\end{align}
and
\begin{align}\label{38}
P_{m,pCSI}^{CD - BH,\infty } \nonumber
=& \sum\limits_{b' \in {\cal B}\backslash \left\{ b \right\}} {{P_{avg}}{{\left( {b'} \right)}_\kappa }} \\\nonumber
 &\times \left\{ {1 - \frac{1}{{K!}}{{\left[ {\frac{{{\tau _m}\left( {{a_{mt}} - {\varepsilon _m}{a_{nt}}} \right)}}{{\chi \left( {{a_{mt}} - {\varepsilon _m}{a_{nt}}} \right)}}} \right]}^K}} \right\} ,\tag{34}
\end{align}
respectively.
\end{corollary}
\begin{remark}
Upon substituting (33) and (34) into (28), the diversity orders of the user m with ipCSI/pCSI for CD-NOMA-BH system are 0 and K, respectively.
\end{remark}

\begin{corollary}\label{Corollary:PD-NOMA:the COP of near user for case1}
For the special case in (26) and (27) $K=1$, the asymptotic outage probability of user m with ipCSI/pCSI at high SNR for PD-NOMA-BH system can be given by
\begin{align}\label{39}
P_{m,ipCSI}^{PD - BH,\infty } \nonumber
=& \left[ {1 - \frac{{{\omega ^*}{\varepsilon _m} + {\tau _m}\left( {{a_{mt}} - {\varepsilon _m}{a_{nt}}} \right)}}{{\chi \left( {{a_{mt}} - {\varepsilon _m}{a_{nt}}} \right)}}} \right]\\\nonumber
& \times \sum\limits_{b' \in {\cal B}\backslash \left\{ b \right\}} {{P_{avg}}{{\left( {b'} \right)}_\kappa }} ,\tag{35}
\end{align}
and
\begin{align}\label{40}
P_{m,pCSI}^{PD - BH,\infty }\nonumber
 = & \left[ {1 - \frac{{{\tau _m}\left( {{a_{mt}} - {\varepsilon _m}{a_{nt}}} \right)}}{{\chi \left( {{a_{mt}} - {\varepsilon _m}{a_{nt}}} \right)}}} \right]  \sum\limits_{b' \in {\cal B}\backslash \left\{ b \right\}} {{P_{avg}}{{\left( {b'} \right)}_\kappa }}  ,\tag{36}
\end{align}
respectively.
\end{corollary}
\begin{remark}
Upon substituting (35) and (36) into (28), the diversity orders of the user m with ipCSI/pCSI for PD-NOMA-BH system are 0 and 1, respectively.

We can observe the above remarks that the CD-NOMA-BH system have multi-carrier effect and obtain upper diversity order than PD-NOMA-BH system. Furthermore, the constraints of the channel estimation error on the ranking order are analyzed. Consequently, the design of the carrier and the excellent channel estimation capability have a prominent impact on U-NOMA-BH systems.
\end{remark}
The relationship between different conditions and  diversity order is expounded in Table I, where we use "D" to represent diversity order.
\begin{table}[!h]
\begin{center}
\caption{The relationship between different factors for CD/PD-NOMA-BH systems.}
{\tabcolsep9.5pt 
\renewcommand\arraystretch{1} 
\begin{tabular}{|c|c|c|c|}\hline   
  \textbf{Pairing users} & \textbf{NOMA}  & \textbf{CSI} & \textbf{D} \\
     \hline
\multirow{4}{*}{The user $m$}    & \multirow{2}{*}{CD-NOMA-BH}   & ipCSI & $0$ \\
\cline{3-4}
                     &    &  pCSI &  $K$ \\
\cline{2-4}
                     & \multirow{2}{*}{PD-NOMA-BH} & ipCSI      &  $0$ \\
\cline{3-4}
                &          & pCSI   & $1$ \\
\hline
\multirow{4}{*}{The user $n$}  & \multirow{2}{*}{CD-NOMA-BH}    &  ipCSI    & $ 0 $ \\
\cline{3-4}
                                  &                                                 &     pCSI   & $ K   $ \\
\cline{2-4}
               &\multirow{2}{*}{PD-NOMA-BH}     &  ipCSI       & $0$  \\
\cline{3-4}
                                  &                                                  &     pCSI   & $ 1   $ \\
\hline
\end{tabular}}{}
\label{Summarize}
\end{center}
\end{table}
\subsection{Throughput Analyses}
The satellite transmits information at a constant rate, and the system throughput is affected by the outage probability.  According to the analytical results derived in the above subsection, using (16), (19), (23) and (25) the CD-NOMA-BH system throughput in the delay-limited mode is expressed as
\begin{align}\label{express p5 }
R_{TH}^{CD} = \left( {1 - P_m^{CD - BH}} \right){R_m} + \left( {1 - P_n^{CD - BH}} \right){R_n}.\tag{37}
\end{align}

Similar to (37), using (20), (21), (26) and (27), the system throughput of PD-NOMA-BH system is given by
\begin{align}\label{express p5 }
R_{TH}^{PD} = \left( {1 - P_m^{PD - BH}} \right){R_m} + \left( {1 - P_n^{PD - BH}} \right){R_n}.\tag{38}
\end{align}
\section{NUMERICAL RESULTS}
In this section, We focus on the numerical analysis of the proposed CD/PD-NOMA-BH systems.  In section V-A, the satellite is designated to cover the equatorial ground region.  The LEO coverage and users demand data are based on\cite{Wang2022Joint,2022Fuzzy}. The satellite parameters and the number of beams refer to 3GPP TR 38.811\cite{3GPP}. Note that the objective value$\sum\limits_{\varphi \in ({\cal M},{\cal N})} {{{\left( {{R_\varphi } - {D_\varphi }} \right)}^ 2}} $ remains consistent with the minimal problem of the objective function in (7a). The number of simulation results exceeds 1000, where per user traffic demand is randomly distributed for each simulation.
In section V-B, we utilize Monte Carlo simulation to discuss the results of performance analysis for the CD/PD-NOMA-BH systems. Moreover, BPCU is used to denote bit per channel use. Additionally, the conventional OMA is regarded as a benchmark for comparison object.  The parameter definitions are summarized in TABLE II unless stated otherwise.

\begin{table}[!t]
\centering
\renewcommand{\thetable}{II}
\caption{Simulation parameters}
\tabcolsep8pt
\renewcommand\arraystretch{1.4} 
\begin{tabular}{|l|l|}
\hline
Parameter  &  Value \\
\hline
Frequency, $f^{\mathrm{fr}}$  & 11.7 GHz   \\
\hline
Bandwidth, $W$  &  200 MHz \\
\hline
Satellite covers the ground longitude range  &  [$85^{\circ} \mathrm{E}$ , $115^{\circ} \mathrm{E}$ ] \\
\hline
Satellite covers the ground Latitude range  &  [$-15^{\circ} \mathrm{S}$ , $15^{\circ} \mathrm{N}$ ] \\
\hline
Satellite location   & $101^{\circ} \mathrm{E}$  \\
\hline
Satellite altitude    &  1000 km \\
\hline
Power budget per user-pair, ${P_s}$    &  5 dBW  \\
\hline
User receive antenna gain, $G_{r}$  & 42.1 dBi  \\
\hline
Satellite transmit antenna gain, $G_{t}$   &  49.6 dBi  \\
\hline
Number of time slots, $T$  & 32  \\
\hline
Number of beams, $B$   &  48  \\
\hline
Maximum active beams, $B_{0}$  &  8,6  \\
\hline
Number of users per beam & 8  \\
\hline
Noise power, $\sigma^2$ & -145 dBW  \\
\hline
Traffic demand, $D_{\varphi}$ & 200 Mbps to 1.4 Gbps \\
\hline
Minimum capacity, $R_\varphi ^{\min }$ & 5 Mbps  \\
\hline
\end{tabular}
\label{parameter}
\end{table}

\subsection{Simulation Results of Optimization}
In this subsection, we compare U-NOMA-BH with three other BH algorithms as follows,

\begin{enumerate}
\item \emph{Orthogonal multiple access beam hopping (OMA-BH): The OMA-BH scheme solves $\mathcal{P}_{0}$ by Algorithm 2. Since there is no need for power and subcarriers allocation, (7b) and (7c) only are considered, and a single user of the beam is analyzed.}
\item \emph{Maximum SINR beam hopping (Max-SINR-BH ): The scheme divides the optimization function into sub-problems in units of each time slot, and decides the allocation of time slots according to the users maximum SINR\cite{Lei2011}. The Algorithm 1 is still used for optimization at user level.}
\item \emph{Periodic beam hopping (P-BH): The scheme adopts round-robin scheduling of beams\cite{Alegre2012}. Each beam is illuminated in turn with the same optimizations for subcarriers and power.}
\end{enumerate}

Fig. 2 plots the performance comparison between proposed U-NOMA-BH systems and the existing OMA-BH, Max-SINR-BH and P-BH in the literature. As can be observed that the U-NOMA-BH systems outperform the other three benchmarks in controlling the difference between user achievable rate and traffic demand. When the average traffic demand is relatively small (less than 450 Mbps), the performance gap between the OMA-BH system and the proposed U-NOMA-BH systems is not widen. However, the difference multiple between U-NOMA-BH and OMA-BH promoting from 0.03 to 0.34 when the demand increases from 450 Mbps to 600 Mbps. This is because that the systems of U-NOMA-BH optimize the subcarriers and power in a single time slot. Compared with Max-SINR-BH and P-BH two BH algorithms, the U-NOMA-BH systems have obvious advantages, which also reduces the error of resource allocation caused by time slots allocation. Additionally, the combination of the unified framework NOMA and  BH can effectively improve the users satisfaction with resource allocation.

\begin{figure}[h!]
    \begin{center}
        \includegraphics[width=3.8in,  height=2.8in]{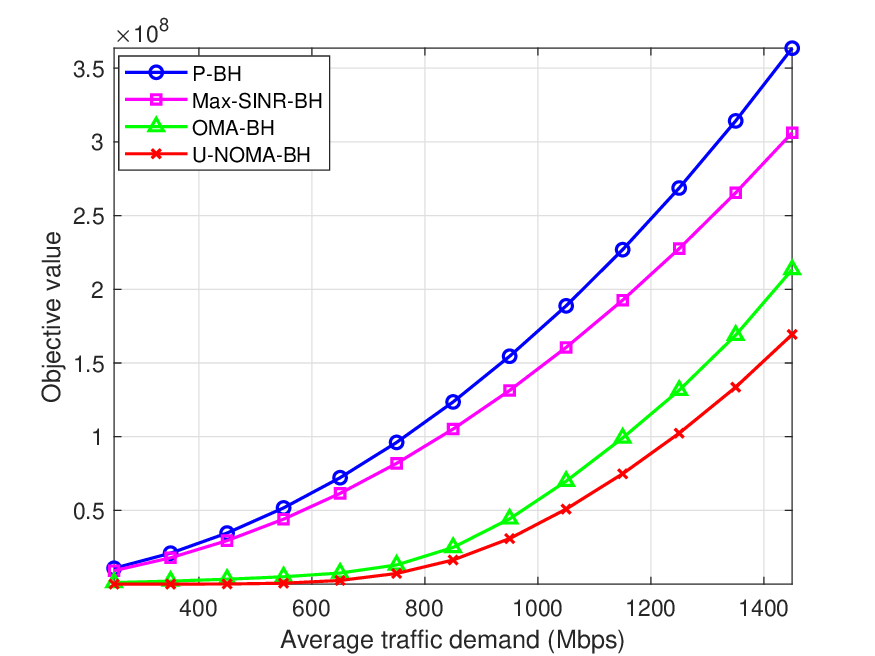}
        \caption{The U-NOMA-BH scheme is compared with the benchmarks in terms of capacity gap.}
        \label{The_SOP_EE_diff_IR}
    \end{center}
\end{figure}

\begin{figure}[h!]
    \begin{center}
        \includegraphics[width=3.8in,  height=2.8in]{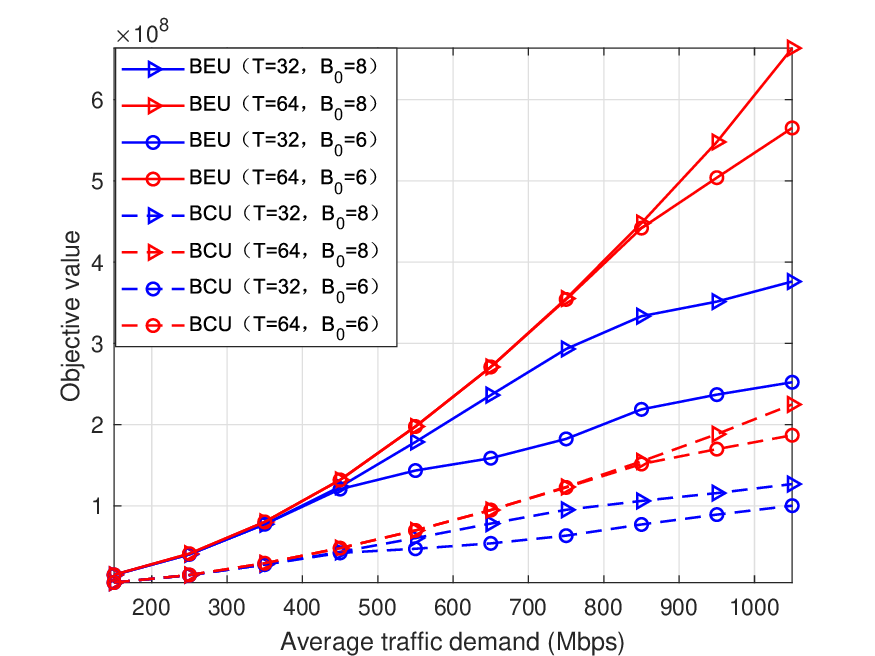}
        \caption{Performance comparison of BCU and BEU based on different timeslots and number of active beams.}
        \label{The_SOP_EE_diff_IR}
    \end{center}
\end{figure}

Additionally, Fig. 3 plots the system of substituting the objective value with BCU and BEU respectively in different cases. To investigate the impact of U-NOMA-BH systems on user fairness, we transform the objective functions into $\mathop {\min }\limits_{{\delta _{bt}},{\beta _{mk}},{a_{mt}}} \sum\limits_{m \in {\cal M}} {{{\left( {{R_m} - {D_m}} \right)}^2}} $ and $\mathop {\min }\limits_{{\delta _{bt}},{\beta _{nk}},{a_{nt}}} \sum\limits_{n \in {\cal N}} {{{\left( {{R_n} - {D_n}} \right)}^2}} $. The performance metrics of BCU and BEU are analyzed by the impact of changes in the number of active beams and time slots, where $T$=32 or 64, and $B_{0}$=8 or 6. We observe that changeing in time slots and active beams number have an impact on the objective values for BEU at 350 Mbps, while BCU is at 450 Mbps. At the same time, the fairness between BCU and BEU are improved by reducing the number of time slots and active beams. The phenomenon indicates that the achieved traffic demand is greater than the constraint objective value.
\begin{figure}[h!]
    \begin{center}
        \includegraphics[width=3.8in,  height=2.8in]{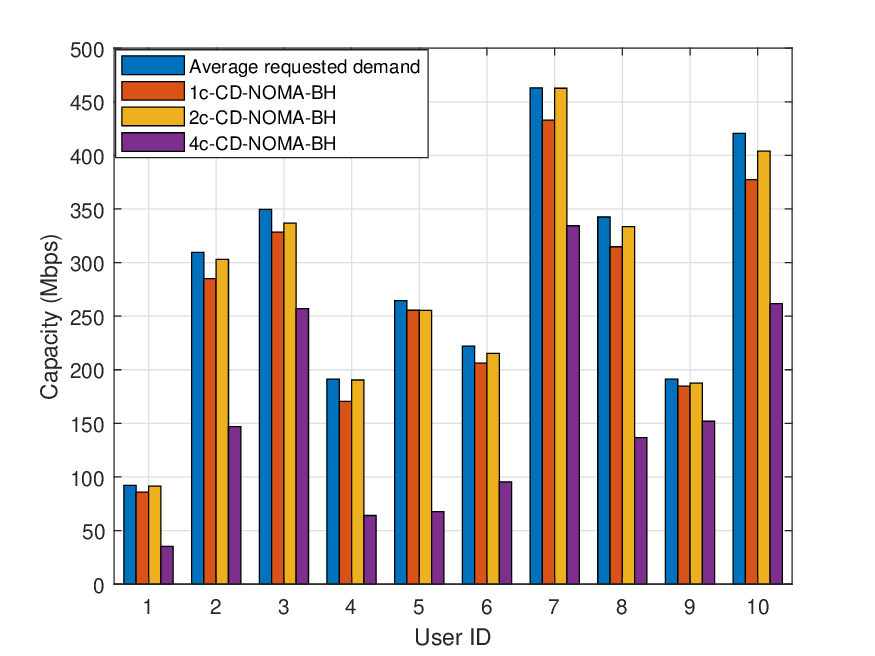}
        \includegraphics[width=3.8in,  height=2.8in]{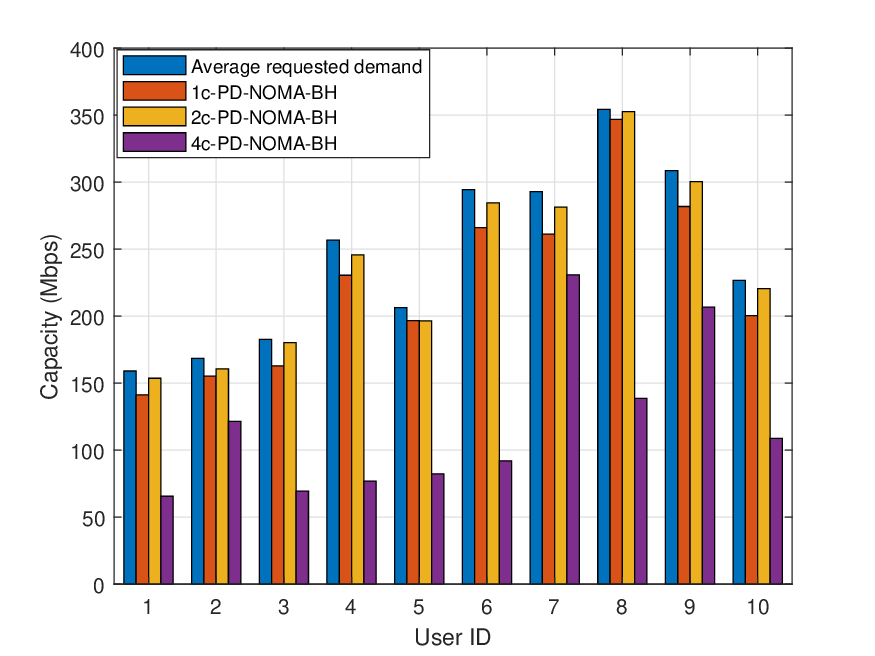}
        \caption{A research of the achievable capacity of users with three different polarization reuse methods and users traffic requirements.}
        \label{The_SOP_EE_diff_IR}
    \end{center}
\end{figure}

We take the case of introducing polarization as different benchmark schemes as follows,

\begin{enumerate}
\item \emph{1c-U-NOMA-BH : The U-NOMA-BH systems are considered for full-band reuse, and the users of active beams in each time slot share a frequency band.}
\item \emph{2c-U-NOMA-BH : The 2-color scheme introduces right and left hand circular polarization, which can reduce the co-channel interference between active beams in the same time slot. The U-NOMA-BH adopts the 2-color scheme to reduce inter-beams interference while ensuring that users have the full bandwidth of the satellite.}
\item \emph{4c-U-NOMA-BH : The 4-color scheme introduces frequency multiplexing on the basis of polarization multiplexing. The combination of U-NOMA-BH and the 4-color scheme makes the inter-beams interference smaller, but also pays the price of sacrificing the frequency band.}
\end{enumerate}

Fig. 4 compares the achievable capacity of the U-NOMA-BH systems introduced by polarization reuse. The bar graph reflect the comparison of U-NOMA-BH combined with the three polarization modes and the user traffic demand. Compared with other schemes, 2c-U-NOMA-BH can better fit the gap between the user achievable capacity and traffic demand. The reason for the phenomenon is that 2-color reuse not only reduces the interference between beams, but also enables users to achieve the full bandwidth, which fulfills the high-traffic requirements of users. More particularly, the 2c-CD-NOMA-BH the performance of users 1 and 7 is outstanding. It shows that the satisfaction of users is reached. The effect of users 4, 6 and 8 are almost a cliff-like difference in 4c-PD-NOMA-BH. This shows that there are gaps between different polarization methods in achieving the satisfaction of users.

\begin{figure}[h!]
    \begin{center}
        \includegraphics[width=3.8in,  height=2.8in]{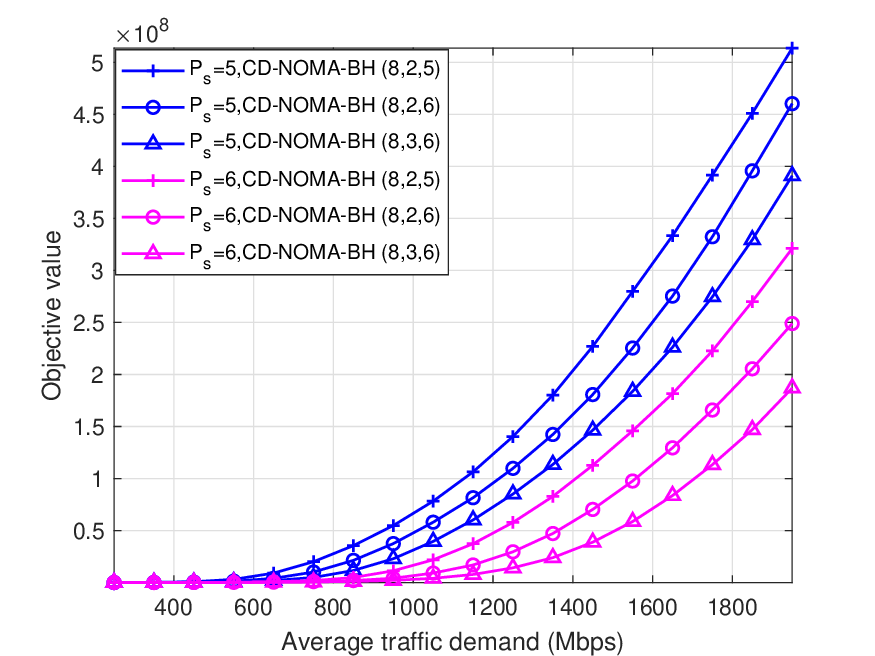}
        \caption{The comparison of objective value performance of CD-NOMA-BH with different carrier conditions.}
        \label{The_SOP_EE_diff_IR}
    \end{center}
\end{figure}

As a further advance, Fig. 5 plots the applicability of the proposed CD-NOMA-BH system for different scenarios. We can observe that the transmit power is increased, the gap between the capacity and the traffic request becomes smaller, which is more suitable for user satisfaction. At the same time, it can be clearly seen from the figure that the subcarrier allocation ratio $(K/Q)$ can significantly improve the performance of the CD-NOMA-BH system with different comparisons  $\left( {M,Q,K} \right)$. In addition, the result shows that the performance of the CD-NOMA-BH system is superior as the number of assignable subcarriers $Q$ grows.
This means that the additional intra-beam interference has less impact on the users achievable capacity, when the CD-NOMA-BH system can accommodate more active subcarriers.
\begin{figure}[h!]
    \begin{center}
        \includegraphics[width=3.8in,  height=2.8in]{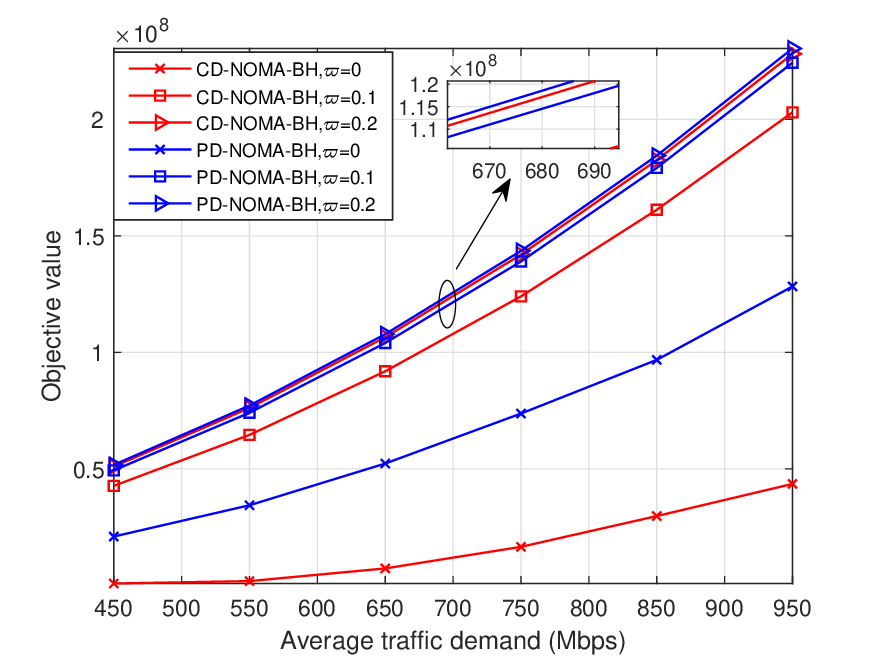}
        \caption{The objective value versus the average traffic demand of users with ipCSI/CSI for CD/PD-NOMA-BH systems.}
        \label{The_SOP_EE_diff_IR}
    \end{center}
\end{figure}

Fig. 6 plots the applicability of the proposed CD/PD-NOMA-BH systems for different channel estimation errors. The red and blue lines represent CD-NOMA-BH and PD-NOMA-BH systems with ipCSI/pCSI, respectively, where $\varpi^*=0.1$, $\varpi^*=0.2$ are postulated. We observe that the severe performance deteriorates as the channel estimation error increasing. This is due to that ipCSI will lead to an overall addition in users interference and decrease the ability to satisfy traffic demands. Hence it is crucial to consider the channel estimation error in practical scenarios for the U-NOMA-BH systems.
\subsection{Simulation Results of Performance Analysis}
Fig. 7 plots the outage probability versus SNR with different channel error and time slots overlap parameters for CD-NOMA-BH system. Based on (23) and (25), the CD-NOMA-BH curves are plotted for different time slots overlap with respect to ipCSI, where ${\varpi ^*} = 0.1$, $\kappa=0.2 \mathrm{~T}$, $\kappa=0.6 \mathrm{~T}$ are set. The optimal power allocations ${a_n} = 0.26$ and ${a_m} = 0.74$ are given according to the optimization algorithm. Apparently, as the number of time slots overlaps increasing, the interruption behavior of paired users becomes worse. This is due to the fact that the variable of time slots overlap mainly comes from the influence of inter-beams interference. A larger amount of time slots overlap will lead to increase the uncertainty of inter-beams interference. For the purpose of comparison, the outage probability curve with pCSI, i.e., ${\varpi ^*} = 0$, is adopted as the benchmark for comparison. It can be observed that the outage probability of user $n$ and $m$ decrease as $\varpi^*$ increases. As a consequence, the practical design of paired NOMA users within a beam should take into account the channel estimation error and the time slots overlap inter-beams.
\begin{figure}[h!]
    \begin{center}
        \includegraphics[width=3.8in,  height=2.8in]{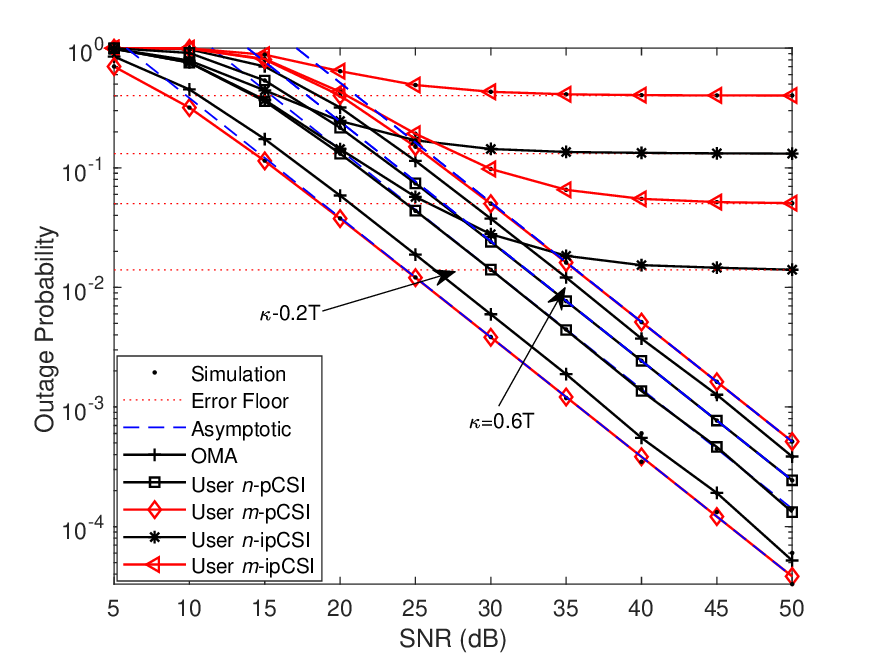}
        \caption{The outage probability of CD-NONA-BH system versus SNR, with $R_{n}=1 \enspace\mathrm{BPCU} $ and $ R_{m}=1.5 \enspace\mathrm{BPCU}$.}
        \label{The_SOP_EE_diff_IR}
    \end{center}
\end{figure}
\begin{figure}[h!]
    \begin{center}
        \includegraphics[width=3.8in,  height=2.8in]{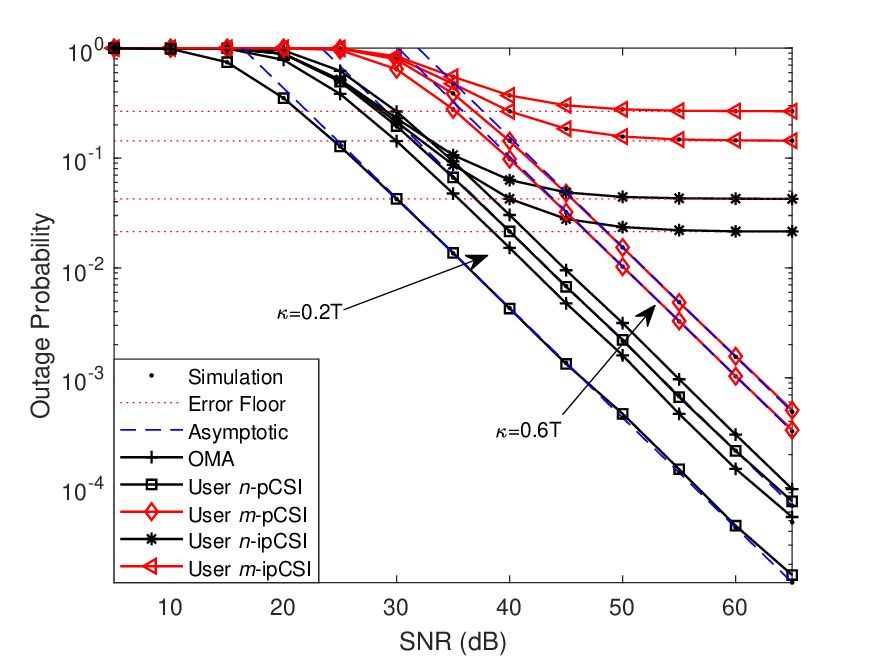}
        \caption{The outage probability of PD-NOMA-BH system versus SNR, with $R_{n}=1 \enspace\mathrm{BPCU} $ and $ R_{m}=1.5 \enspace\mathrm{BPCU}$.}
        \label{The_SOP_EE_diff_IR}
    \end{center}
\end{figure}

Fig. 8 plots the effect of channel estimation error in the case of PD-NOMA-BH system on the outage performance versus transmission SNR. For the special case of $K = 1$, the outage probability curves of the BCU and BEU for PD-NOMA-BH system are plotted according to (20), (21), (26) and (27). Furthermore, Monte Carlo simulation curves of BCU and BEU are plotted for ipCSI/pCSI with different amount of time slots overlap and coincide with the derived results. It can be observed that the outage probability of PD-NOMA-BH system is higher than that of CD-NOMA-BH. The reason for this phenomenon is that multiple subcarriers can reduce interference within the beam.
\begin{figure}[h!]
    \begin{center}
        \includegraphics[width=3.8in,  height=2.8in]{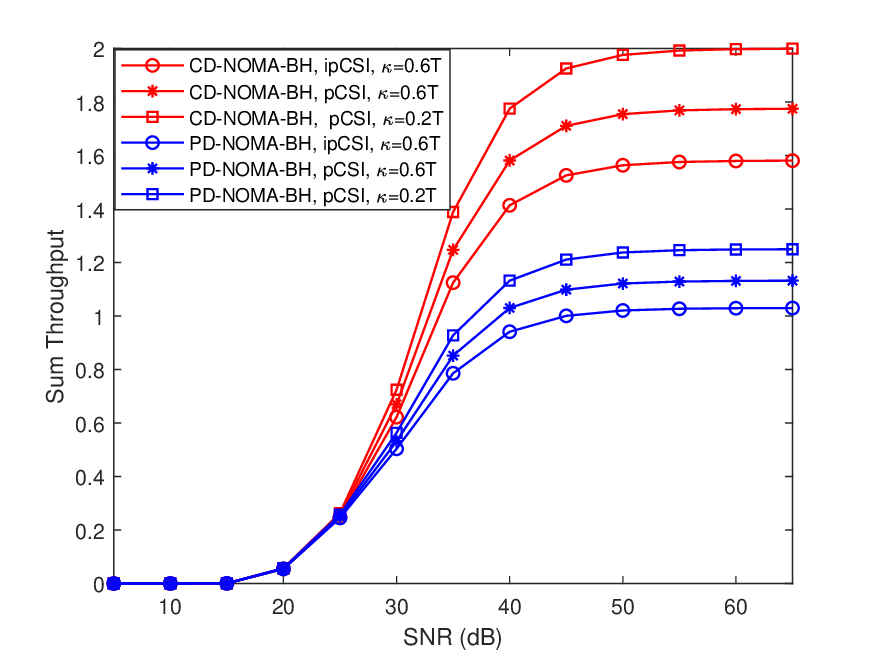}
        \caption{System throughput of CD/PD-NOMA-BH systems versus SNR with ipCSI/pCSI.}
        \label{The_SOP_EE_diff_IR}
    \end{center}
\end{figure}

Fig. 9 plots the effect of the considered channel error and time slots overlap on the system throughput. The red and blue lines represent the system throughput of CD-NOMA-BH and PD-NOMA-BH, respectively, which is obtained by (37) and (38). As can be observed that the system throughput of CD-NOMA-BH is better than that of PD-NOMA-BH. This is because that CD-NOMA-BH achieves the smaller outage probabilities. One can observed that the system throughput still decrease with increasing time slots overlap. The system throughput of ipCSI that we focus on is weaker than that of pCSI.

\section{Conclusion}
In this paper, the U-NOMA-BH systems are proposed, where contains an optimized CD/PD-NOMA-BH and the performance of U-NOMA-BH systems. We propose a resource allocation problem to minimize the gap between user traffic demand and achieved capacity by jointly optimizing power allocation, beam level time slots, and subcarriers allocation. The results specified that the CD/PD-NOMA-BH algorithm designed is better than P-BH, OMA-BH and Max-SINR-BH. In addition,  the closed expressions of the outage probabilities of BCU and BEU have been derived based on the above optimization algorithm. According to the analysis results, it has been demonstrated that the outage probability of CD-NOMA-BH is better than that of PD-NOMA-BH. Finally, the system throughput in delay-limited modes has been studied. Numerical results verified the outcome of the analysis. The diverse scenarios and challenging tasks applying in U-NOMA-BH systems are able to further provide additional design options，which is a promising future research direction.
\bibliographystyle{IEEEtran}
\bibliography{mybib}

\end{document}